\documentclass{draft}
\usepackage{graphicx}
\usepackage{overpic}
\usepackage{epstopdf, epsfig}
\usepackage{color}
\usepackage{amsmath}
\usepackage{amsfonts}
\usepackage{upgreek}
\usepackage{amssymb}
\usepackage{multirow}
\usepackage{float}
\usepackage{verbatim}
\usepackage{appendix}

\shorttitle{Drag reduction by inertialess spheroids}
\shortauthor{Z. Wang, C.-X. Xu and L.-H. Zhao}

\title{Turbulence modulations and drag reduction by inertialess spheroids in turbulent channel flow}

\author{Ze Wang\aff{1}
  Chun-Xiao Xu\aff{1}
 \and Lihao Zhao\aff{1}
\corresp{\email{zhaolihao@tsinghua.edu.cn}},
}

\affiliation{\aff{1}AML, Department of Engineering Mechanics, Tsinghua University, 100084 Beijing, China}

\begin{document}

\maketitle

\begin{abstract}
Previous studies on nonspherical particle-fluid interaction were mostly confined to elongated fiber-like particles, which were observed to induce turbulence drag reduction. However, with the presence of tiny disk-like particles how wall turbulence is modulated and whether drag reduction occurs are still unknown. Motivated by those open questions, we performed two-way coupled direct numerical simulations of inertialess spheroids in turbulent channel flow by an Eulerian-Lagrangian approach. The additional stress accounts for the feedback from inertialess spheroids on the fluid phase. The results demonstrate that both rigid elongated fibers (prolate spheroids) and thin disks (oblate spheroids) can lead to significant turbulence modulations and drag reduction. However, the disk-induced drag reduction is less pronounced than that of rigid fibers with the same volume fraction. Typical features of drag-reduced flows by additives are observed in both flow statistics and turbulence coherent structures. Moreover, in contrast to one-way simulations, the two-way coupled results of spheroidal particles exhibit stronger preferential alignments and lower rotation rates. At the end we propose a drag reduction mechanism by inertialess spheroids and explain the different performance for drag reduction by fibers and disks. We find that the spheroidal particles weaken the quasistreamwise vortices through negative work and, therefore, the Reynolds shear stress is reduced. However, the mean shear stress generated by particles, which is shape-dependent, partly compensates for the reduction of Reynolds shear stress and thus affects the efficiency of drag reduction. The present study implies that tiny disk-like particles can be an alternative drag reduction agent in wall turbulence.
\end{abstract}


\section{Introduction}
 Drag reduction in wall-bounded turbulence is of importance in industrial applications \citep{kim_physics_2011}. Toms  \citep{Toms_Some_1949} observed a dramatic drag reduction in turbulent flows induced by adding a small amount of long-chain flexible polymer and followed by Toms's work the polymer induced by drag reduction has been widely investigated \citep{white_mechanics_2008,benzi_polymers_2018}. Compared with polymer additives, tiny rigid fiber-like particles are also found to produce drag reduction in wall-bounded turbulence but less efficient \citep{radin_drag_1975,gyr_drag_1995}. However, the rigid particles are more tolerant with respect to shear-induced degradation \citep{reddy_drag_1985}. Therefore, rigid fibers are of significance to sustain the drag reduction in wall-bounded turbulence. 
 
The small rigid fibers are often approximated as inertialess and elongated prolate spheroids in theoretical and numerical investigations (e.g. \cite{brenner_rheology_1974,gillissen_fibre-induced_2008}). The spheroids with small size influence the surrounding flow via additional stress and the feedback forces and torques are often neglected with negligibly small particle inertia, i.e. inertialess \citep{guazzelli_physical_2011}. Batchelor's pioneering theoretical work on micro-hydrodynamics \citep{batchelor_slender-body_1970,batchelor_stress_1970,batchelor_stress_1971} advanced the understanding of particle suspensions. He firstly derived the expression of bulk stresses in a dilute suspension of force-free ellipsoidal particles \citep{batchelor_stress_1970}. The particle stress generated by elongated prolate spheroids in a dilute or semi-dilute suspensions is also described based on slender-body theory \citep{batchelor_slender-body_1970,batchelor_stress_1971}. \cite{brenner_rheology_1974} developed a general rheology theory for a dilute suspension of rigid axisymmetric particles. Several studies \citep{doi1986The,hinch_effect_1972,hinch_constitutive_1976,shaqfeh_hydrodynamic_1990} further investigated the particle stress in a suspension of elongated spheroids theoretically. On the other hand, direct numerical simulations (DNSs) of suspensions of inertialess spheroids have not been performed until 1990s to computational limitations. With an aligned-particle approximation \citep{lipscomb_flow_1988}, direct numerical simulations preformed by \cite{den_toonder_drag_1997} showed that elongated spheroids lead to a significant drag reduction and the modulated turbulence statistics qualitatively agreed with the experimental measurements in turbulent pipe flows.  \cite{paschkewitz_numerical_2004} performed a two-way coupled DNS of suspensions of rigid fibers in a minimal channel flow \citep{jimenez_minimal_1991}. They demonstrated that the fluctuations of particle stress in inter-vortex extensional regions weaken the near-wall vortex structures and induce turbulent drag reduction. \cite{gillissen_fibre-induced_2008}  carried out coupled Eulerian simulations and observed that tiny elongated spheroids lead to a reduction of the frictional drag in wall turbulence at various Reynolds numbers. A two-way coupled Eulerian-Lagrangian approach is adopted by \cite{moosaie_direct_2013} to investigate drag reduction caused by rigid fibers. This Lagrangian point-particle tracking method is computationally expensive, because the particle orientation is solved directly without using any closure model or artificial diffusivity  \citep{gillissen_stress_2007}.

The suspensions of rigid fibers, modeled as prolate spheroids, attracted more attentions in the past decade. However, the case of disk-like particles, which can be modeled as oblate spheroids, is also commonly encountered in environmental, biological and industrial processes, for instance, clay minerals in water \citep{whalley_oriented_1992,ruzicka_fresh_2011}, erythrocytes in blood \citep{shardt_direct_2012} and graphene oxide suspensions \citep{dan_liquid_2011}. Nevertheless, only few studies were carried out on the turbulent flows laden with tiny disk-like particles. \cite{gust_observations_1976} experimentally reported the occurrence of turbulent drag reduction in a clay-mineral suspension, but he ascribed that to the agglomeration of particles. \cite{tesfai_rheology_2013} observed a very high intrinsic viscosity in a dilute suspension of graphene oxide. Recent studies \citep{wang_particle_2020,moosaie_rheology_2020} showed that thin oblate spheroids produce strong fluctuations of particle shear stresses in one-way coupled direct numerical simulations. Oblate spheroids are thus expected to have non-negligible influences on turbulent fluid phase. However, due to the lack of two-way coupled high-fidelity simulations and convincing experiments, how turbulent flows are modulated with the presence of tiny oblate spheroids and whether drag reduction can be induced still remain open questions. 

In the present work, we perform two-way coupled DNS and focus on the interaction between inertialess spheroidal particles and fluid turbulence via particle stress in a channel flow. The tiny disks and fibers are modeled as inertialess oblate and prolate spheroids, respectively. We demonstrate for the first time that adding thin disk-like particles can lead to significant turbulence  modulations and a pronounced drag reduction, which is, however, less pronounced than that of rigid fibers.  The remainder of this paper is organized as follows. In the section  of Methodology, the governing equations for a dilute suspension of spheroids and the detailed numerical methods are described. Section 3.1 and 3.2 present the statistics of  fluid and particle, respectively. A mechanism for drag reduction by inertialess spheroids is proposed in Section 3.3. Conclusions are drawn in Section 4.

\section{Methodology}\label{methodology}

\subsection{Governing equations}\label{equations}

\subsubsection{Eulerian fluid phase}\label{fluid phase}
The turbulent channel flow of a dilute suspension is computed by DNS in an Eulerian framework. The carrier fluid is incompressible, isothermal, and Newtonian. The fluid motion is governed by mass continuity and momentum equations:

\begin{gather}
\frac{\partial u_{i}}{\partial x_{i}}=0, \\
\rho\left(\frac{\partial u_{i}}{\partial t}+ u_{j}\frac{\partial u_{i}}{\partial x_{j}}\right)=-\frac{\partial p}{\partial x_{i}}+\mu \frac{\partial^{2} u_{i}}{\partial x_{j} \partial x_{j}}+\frac{\partial \tau_{i j}^{p}}{\partial x_{j}} .
\end{gather}
where $p$ is the pressure, $\rho$ and $\mu$ are the fluid density and dynamic viscosity, respectively. The particle stress tensor $\tau_{i j}^{p}$ represents the influence of inertialess spheroids on the fluid. A friction Reynolds number is defined as $Re_{\tau} = u_{\tau}h/\nu$, based on the channel half-height $h$ and the nominal friction velocity
\begin{equation}
  u_{\tau}= \sqrt{-\frac{h}{\rho} \frac{\mathrm{d}\langle p_w\rangle}{\mathrm{d}x}}=\sqrt{\frac{\tau_w}{\rho}},
 \end{equation}
where ${\mathrm{d}\langle p_w \rangle}/{\mathrm{d}x}$ is the mean pressure gradient in the streamwise direction and $\tau_w=-h{\mathrm{d}\langle p_w \rangle}/{\mathrm{d}x}$ is the wall shear stress. Hereafter, the superscript $'+'$ denotes the normalization by viscous scales for velocity ($u_{\tau}$), length ($\nu/u_{\tau}$), and time ($\nu/{u_\tau}^2$).

\subsubsection{Lagrangian particle phase}\label{particle phase}
Lagrangian point-particle approach is adopted to describe the dynamics of small  inertialess particles. These non-Brownian tracers passively follow the translational motion of local fluid, but the rotational motion is determined by the following equations \citep{jeffery_motion_1922} :

\begin{gather}
\omega_{x^{\prime}}=-\kappa E_{y^{\prime} z^{\prime}}+R_{x^{\prime}},\notag \\
\omega_{y^{\prime}}=\kappa E_{x^{\prime} z^{\prime}}+R_{y^{\prime}},\notag \\
\omega_{z^{\prime}}=R_{z^{\prime}}.
\end{gather}
Here, $\omega_{i^{\prime}}$ is the particle rotation vector, $E_{i^{\prime}j^{\prime}}$ is the fluid rate-of-strain tensor and $R_{i^{\prime}}$ is the fluid rate-of-rotation tensor, defined in a particle frame of reference $\mathbf{x}_{i^{\prime}} = \langle x^{\prime}, y^{\prime}, z^{\prime}\rangle$. The transformation between the inertial and the particle frame is obtained by an orthogonal matrix \citep{Goldstein1980Classical} . The shape parameter $\kappa = (\lambda^2-1)/(\lambda^2+1)$ denotes the eccentricity, where $\lambda$ is the aspect ratio defined as the ratio between the length of the symmetry axis (in the $z^{\prime}$ direction) and that of the two equal axes. Therefore, prolate spheroids have $\lambda>1$ and $\kappa>0$, while oblate ones have $\lambda<1$ and $\kappa<0$.

\subsubsection{Particle stress}\label{particle stress}
The spheroids considered in the present study are  inertialess and, therefore, force-free and torque-free, and particles affect the fluid flow via particle stress. This additional stress in a dilute suspension corresponds to the local volume average of stresslets, which represent the resistance of rigid particles to a straining motion \citep{batchelor_stress_1970,guazzelli_physical_2011}. \cite{brenner_rheology_1974}  derived the expression of the symmetric stresslet ($S_{ij}$) for a spheroidal particle:
\begin{equation}
S_{i j}=5\mu V_{p} \left [2 Q_{1} E_{i j}+Q_{2} \delta_{i j} E_{k l} n_{k} n_{l}+2 Q_{3}\left(n_{i} n_{l} E_{l j}+E_{i k} n_{k} n_{j}\right)+Q_{4} n_{i} n_{j} n_{k} n_{l} E_{k l} \right ].
\end{equation}
Here, $V_p$ is particle volume. The material constant $Q_{\alpha} (\alpha=1, 2, 3, 4)$ is a function of aspect ratio $\lambda$. A direction cosine $n_{i} (i=x, y, z)$ is defined as the cosine of the angle between the particle symmetry axis and the $x_i$-direction. The particle stress produced by $N_p$ particles within a given volume $\bigtriangleup$ (e.g. a grid cell) is expressed as,
\begin{equation}
\tau_{i j}^{p}=\frac{1}{\bigtriangleup}\sum_{\beta =1}^{N_p}S_{i j}^\beta,
\end{equation}
where $S_{i j}^\beta$ is the stresslet of the $\beta$th particle. The inertialess spheroids are uniformly distributed and a sufficiently large amount of particles is needed to reach the given volume fraction in the whole channel. It is clear that the volume fraction plays a crucial role concerning the interaction between particles and turbulence. The two-way coupled simulations are implemented by substituting the stress tensor $\tau_{i j}^{p}$ into Eq.(2.2). Similar numerical approaches have been employed in earlier studies \citep{terrapon_lagrangian_2005,moosaie_direct_2013}.

\subsection{Simulation setup }\label{simulation setup}
As described in Section 2.1, we use an Eulerian-Lagrangian point-particle approach to simulate the suspensions. The two-way coupling scheme accounted for the feedback from inertialess spheroids on fluid phase via particle stress. 

A fully developed turbulent channel flow at friction Reynolds number $Re_\tau =180$ is considered. Flow is driven by a constant mean pressure gradient in the streamwise direction. Simulations are performed on a $10h \times 5h \times 2h$ channel domain using $128\times 128 \times 128$ grid points in the streameise $x$, spanwise $y$ and the wall-normal $z$ directions, respectively. The grids are uniformly distributed in the homogeneous directions with $\Delta x^+= 14.0$ and $\Delta y^+= 7.0$. The mesh size $\Delta z^+$ varies between $1.3$ and $4.3$ with a refinement toward walls.We impose periodic boundary conditions in the $x$ and $y$ directions and no-slip and impermeability conditions at the channel walls. Spatial derivatives are computed by a pseudo-spectral method in the two homogeneous directions and a second-order central finite-difference method in the wall-normal direction. The governing equations are integrated forward in time by using an explicit second-order Adams-Bashforth scheme with a time step $\Delta t^+ =0.036$.

Inertialess spheroids with aspect ratios of $\lambda =100$, $0.01$ and $0.002$ are considered to explore the effect of particle shape on turbulence modulations. In two-way coupled simulations, 40 million particles of each type are released randomly into fully developed turbulent channel flows. The volume fraction is chosen as $0.75\%$, which is the same as in previous simulations \citep{paschkewitz_numerical_2004,moosaie_direct_2013}. It should be noted that the equivalent diameter of particles is much smaller than the local Kolmogorov length, but the major axes may on the order of that due to the high asphericity. Nevertheless, the tracer model is still applicable since the particle size has a marginal influence on tiny spheroids with small Stokes number \citep{shin_rotational_2005,parsa_inertial_2014,ravnik_application_2018}. Particle-particle collisions are ignored to focus on the turbulence modulations induced by particle stress. A fully elastic collision model is employed for particle-wall collisions. A second-order scheme is adopted to update the particle position and orientation, with a quadratic interpolation scheme to compute the local fluid field. The time step used is equal to that in Navier-Stokes equations. We use a fully developed unladen turbulent flow at $Re_\tau =180$ as the initial field in all simulations. Statistics are gathered over a temporal sampling interval $\Delta T^+ = 3600$ after the bulk flow reaches a steady state. One-way coupled simulations is also conducted for comparison purposes. We follow the same approach as that adopted in our  earlier work \citep{challabotla_shape_2015,zhao_mapping_2019}. One million tiny particles of each shape are immersed in the same turbulent field without considering the influence of particles on fluid phase. The fluid and particle statistics are collected over a time window of $3600<t^+<7200$.

\section{Results and discussion}\label{results and discussion}

\subsection{Fluid statistics}

\begin{table} 
	\begin{center}
		\def~{\hphantom{0}}
		\begin{tabular}{lccc}
			Aspect ratio   & Volume fraction   & Bulk velocity enhancement & Drag reduction \\[3pt]
			$\lambda = 100$   & 0.75\%            & 8.42\%                    & 14.93\%\\
			$\lambda = 0.01$  & 0.75\%            & 0.98\%                    & 1.92\%\\
			$\lambda = 0.002$ & 0.75\%            & 3.78\%                    & 7.15\%\\
		\end{tabular}
		\caption{Drag reduction of channel flows laden with spheroids of different shapes.}
		\label{tab:reduction}
	\end{center}
\end{table}

The turbulent flows for all cases are driven by the same mean pressure gradient in the streamwise direction and the increase in bulk velocity $U_b$ is indicative of drag reduction. The drag coefficient is defined as $C_f=\tau_w/0.5\rho U_b^2$. As showing in Table 1, the degree of drag reduction with the same particle volume fraction ($0.75\%$) is strongly dependent on the shape of spheroids. The prolate spheroids with $\lambda = 100$ lead to an enhancement of $8.42\%$ of the bulk velocity and  $14.93\%$ reduction of the drag coefficient. The results for the suspension of prolate spheroids are in agreement with previous observations \citep{paschkewitz_numerical_2004,moosaie_direct_2013}. It is observed that the oblate spheroidal particles with $\lambda = 0.01$ only lead to a modest drag reduction, which becomes significant for the flattest disks with $\lambda = 0.002$.

Figure 1 shows the normalized mean streamwise velocity profiles for unladen flow (one-way coupling) and particle-laden flows (two-way coupling). From Figure 1(a), the velocity is enhanced for particle-laden flow compared with that for Newtonian fluid in the region away from the wall, which is consistent with the results of Table 1. However, there is an attenuation in the near-wall region in the semi-log plot (figure 1(b)), especially for the suspension of the flattest spheroids. Interestingly, the intersection between the velocity profile of particle-laden flow and that of unladen flow in the buffer layer moves away from the wall as aspect ration of particles decreases. 

\begin{figure} 
	\centering
	\begin{tabular}{ccc}	
		\begin{overpic}[width=6.5cm
			]{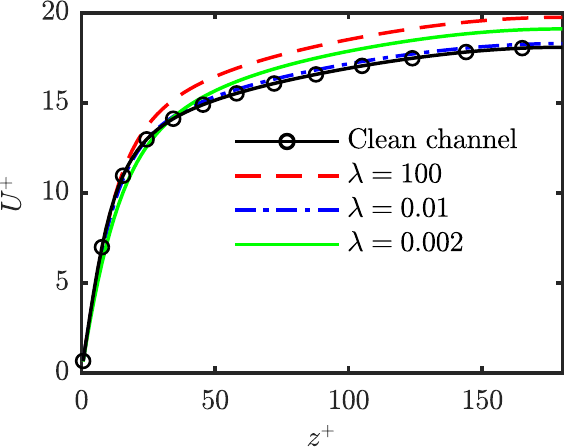}
			\put(0,70){\textbf {(a)}} 
		\end{overpic}  
		&
		\begin{overpic}[width=6.5cm
			]{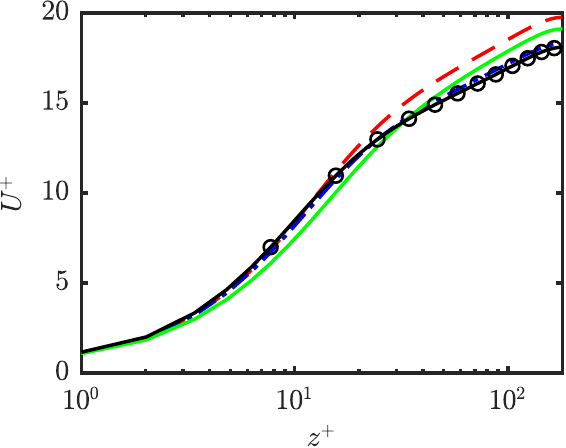}
			\put(0,70){\textbf{(b)}} 
		\end{overpic} \\	   
	\end{tabular}
	\caption{Mean fluid velocity profiles in the streamwise direction. (a) linear plot; (b) semi-log plot.}
	\label{fig1}
\end{figure}

The shear stress balance is derived from the integration of the streamwise mean-momentum equation:
\begin{equation}
\mu\frac{\mathrm{d}U}{\mathrm{d}z}-\rho \langle u'w' \rangle +\langle \tau_{xz}^p \rangle =\tau^{total}= -h\frac{\mathrm{d} \langle p_w \rangle }{\mathrm{d}x}(1-\frac zh)
\label{shear stress balance}.
\end{equation}
 The distribution of the viscous, Reynolds, and particle shear stress in the wall-normal direction is plotted in Figure 2. The linear profile of total shear stress confirms that the flow has reached a statistically steady state and satisfies the integral momentum balance. In comparison with the unladen flow, Reynolds shear stress for particle-laden flow decreases throughout the whole half channel, while viscous shear stress is attenuated in the near-wall region and augmented from the buffer layer. The flattest disks with $\lambda = 0.002$ induce the maximum particle shear stress and have the greatest impact on shear stress balance.
 
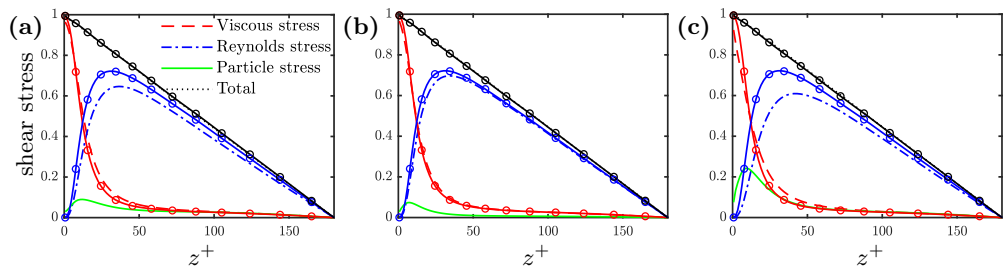
\begin{figure}
    \flushright
	\begin{tabular}{ccc}	
		\begin{overpic}[width=3.9cm
			]{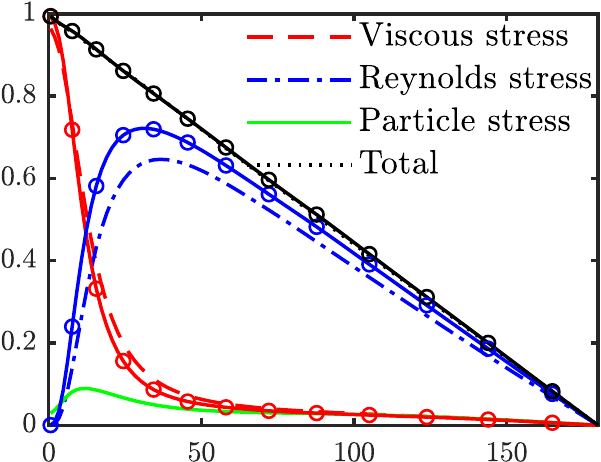}
			\put(-11,68){\textbf{(a)}}
			\put(-11,20){ \rotatebox{90}{shear stress}}			 
			\put(50,-10){\text {$z^+$}}
		\end{overpic}
	    \hspace{0.2cm} 
	  
		&
		\begin{overpic}[width=3.9cm
			]{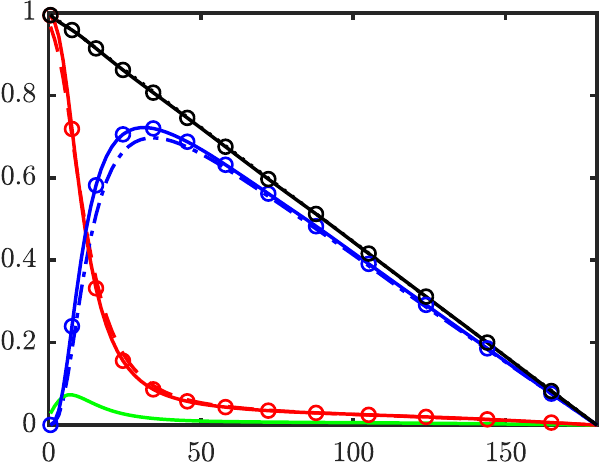}
			\put(-11,68){\textbf{(b)}} 
			\put(50,-10){\text {$z^+$}} 
		\end{overpic} 
     	\hspace{0.2cm} 
      &
		\begin{overpic}[width=3.9cm
			]{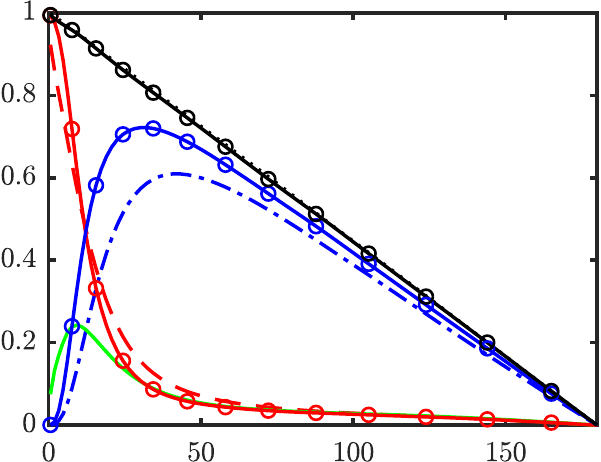}
			\put(-11,68){\textbf{(c)}}  
	    	\put(50,-10){\text {$z^+$}} 
		\end{overpic}   \\
	\end{tabular}
    \vspace{0.4cm}
	\caption{Stress balance for laden and unladen flows (solid lines with circles). (a) $\lambda = 100$; (b) $\lambda = 0.01$; (c) $\lambda =0.002$.}
	\label{fig2}
\end{figure}

As shown in Figure 3, turbulent kinetic energy $k$ and turbulence intensities rms$(u_i')$ are also altered by the presence of inertialess spheroids. The  $k^+$ and rms$(u'^+)$ are enhanced except in the near-wall region, while turbulence intensities in the spanwise and wall-normal directions are attenuated. Additionally, all peaks are shifted further away from the wall compared with that in the unladen flow. The results of intensities imply the modified anisotropy of turbulence in particle-laden flows presented in Figure 4.
The flows with spheroids are less isotropic in the channel center and more prolate axisymmetric in the region away from the wall. The addition of particles also makes the right tip approach the one-component limit. Interestingly, the wall value for the suspension of disk-like particles tends towards the state of isotropic two-component turbulence, contrary to that for the suspension of rod-like particles.
\cite{frohnapfel_interpretation_2007}  concluded that drag-reduced channel flow is commonly accompanied with the increased anisotropy of turbulence in the near-wall region. The modulations of the Lumley anisotropy map indicate that disks  are less effective than rods in drag reduction.

\begin{figure}
	\centering
	\begin{tabular}{ccc}	
		\begin{overpic}[width=6.5cm
			]{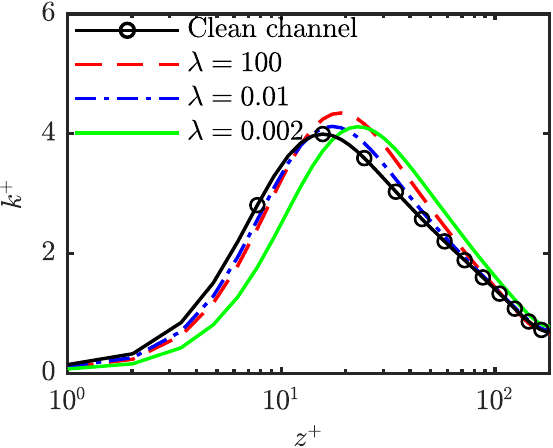}
			\put(0,70){\textbf{(a)}} 
		\end{overpic}  
		&
		\begin{overpic}[width=6.5cm
			]{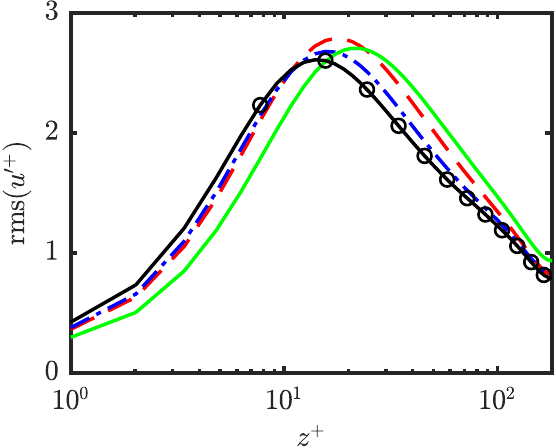}
			\put(0,70){\textbf{(b)}}  
		\end{overpic} \\
		\begin{overpic}[width=6.5cm
			]{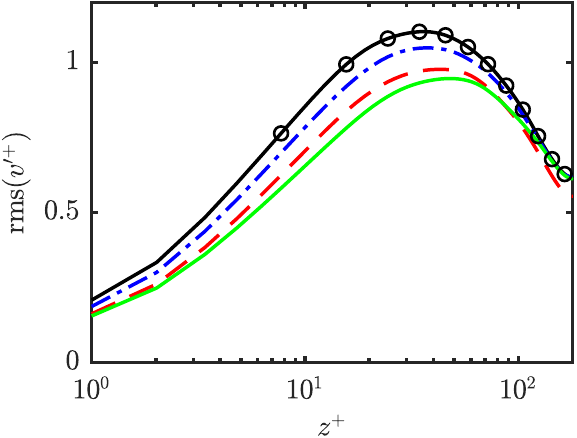}
			\put(0,70){\textbf{(c)}}
		\end{overpic}  
		&
		\begin{overpic}[width=6.5cm
			]{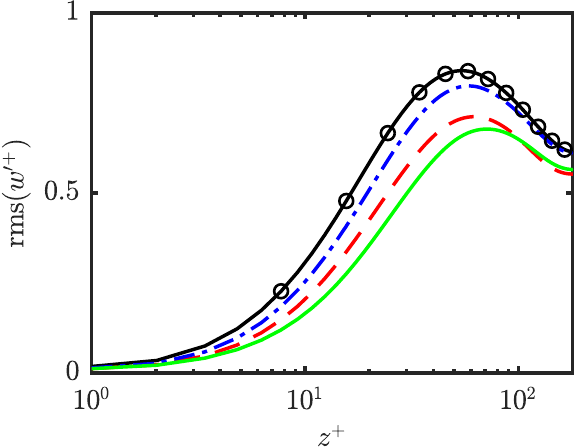}
			\put(0,70){\textbf{(d)}} 
		\end{overpic} \\
	\end{tabular}
	\caption{(a) Turbulent kinetic energy and (b-d) turbulence intensities in the streamwise, spanwise, and wall-normal directions, respectively.}
	\label{fig3}
\end{figure}

\begin{figure}
	\centering
	\begin{overpic}[width=6.5cm
		]{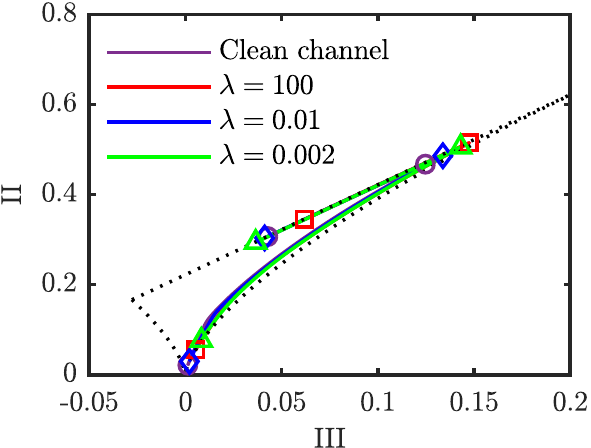}
	\end{overpic} 
	\caption{Lumley anisotropy map \citep{lumley_newman_1977} of the unladen flow and particle-laden flows. The second and third invariants for turbulence are defined as  $\mathrm{\uppercase\expandafter{\romannumeral2}}=a_{ij}a_{ji}$, $\mathrm{\uppercase\expandafter{\romannumeral3}}=a_{ij}a_{jk}a_{ki}$, respectively, where  $a_{ij}= \langle u'_i u'_j\rangle /2k -\delta_{ij}/3$. The dotted line represents the Lumley triangle, which shows the limiting states of turbulence. The states of the wall, right tip, and channel centerline are highlighted by symbols.}
	\label{fig4}
\end{figure}

The statistical results of the vorticity field are depicted in Figure 5. The mean spanwise vorticity corresponds to the mean velocity gradient and resembles the profile of the viscous shear stress in Figure 2. In the near-wall region, the decrease of mean spanwise vorticity caused by spheroids is due to the attenuation of local mean velocity. As shown in Figure 5(b-d), the vorticity fluctuations are damped in the particle-laden channel, indicating the suppression of vortical structures. The distance between the local maximum and minimum of the streamwise fluctuations represents the average radius of the streamwise vortices, while the magnitude of the local maximum reflects the average strength \citep{kim_turbulence_1987}. Therefore, figure 5(b) manifests the increased size and the decreased strength of the streamwise vortices in laden flows as compared to that in unladen flow. Since the spanwise vorticity is mainly contributed by ${\partial u^{\prime}}/{\partial z}$ in the near-wall region, the reduced rms$(\Omega^{'+}_y)$ also indicates that the mean streak spacing is increased due to the presence of spheroids.

\begin{figure}
	\centering
	\begin{tabular}{ccc}	
		\begin{overpic}[width=6.5cm
			]{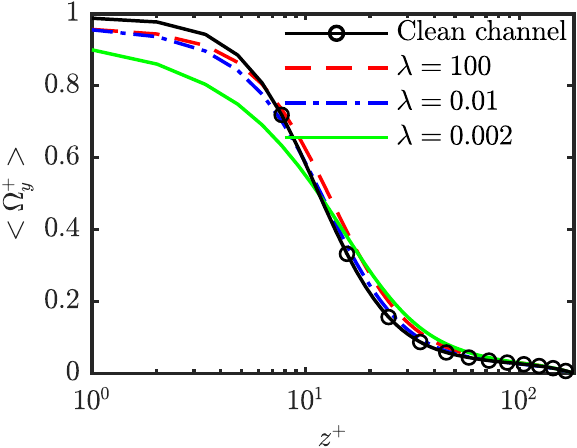}
			\put(0,70){\textbf{(a)}} 
			
		\end{overpic}  
		&
		\begin{overpic}[width=6.5cm
			]{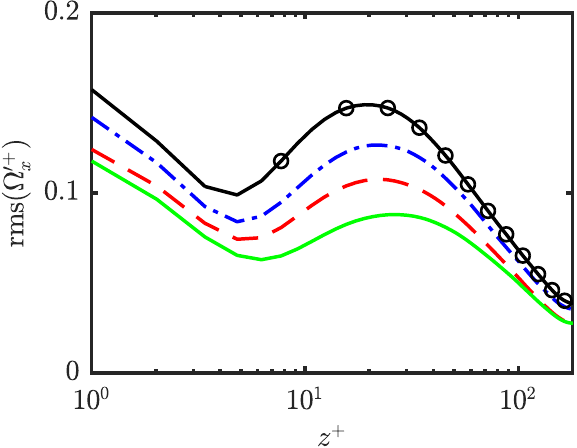}
			\put(0,70){\textbf{(b)}}
		\end{overpic} \\
		\begin{overpic}[width=6.5cm
			]{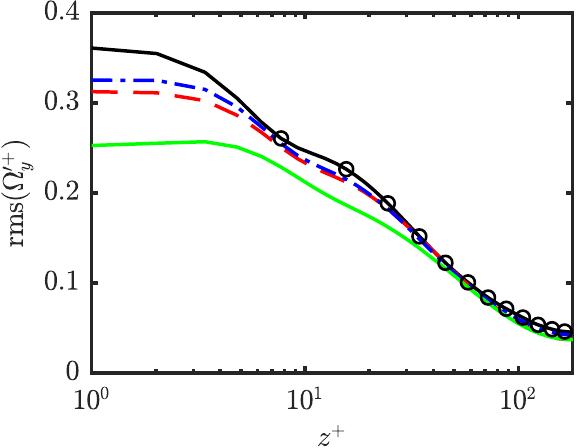}
			\put(0,70){\textbf{(c)}} 
		\end{overpic}  
		&
		\begin{overpic}[width=6.5cm
			]{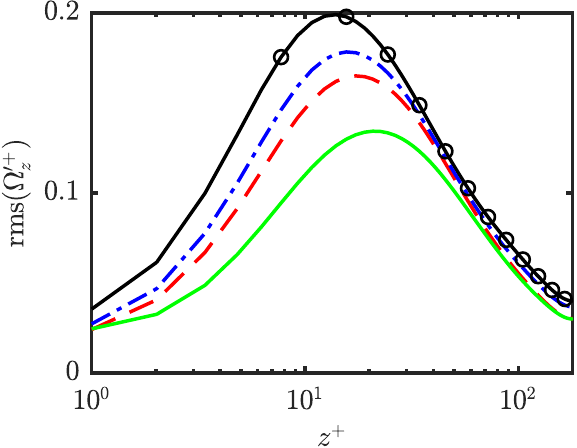}
			\put(0,70){\textbf{(d)}} 
		\end{overpic} \\
	\end{tabular}
	\caption{(a) Mean spanwise vorticity and (b-d) root-mean-square of the streamwise, spanwise and wall-normal fluctuating fluid vorticity components, respectively.}
	\label{fig5}
\end{figure}

\subsection{Particle statistics}

From the previous results, it is known that inertialess spheroids could affect the fluid flows. On the other hand, the modulated fluid field will be influential in the particle dynamics, in return. Since the massless spheroids passively translate along with the local fluid, the focus in this section is on orientational and rotational behavior.

Figure 6 shows the mean absolute values of the direction cosines in particle-laden flows compared with that in one-way coupled simulations. We observe that disks tend to align in the wall-normal direction, while rods preferentially align with the wall. The shape-dependence of particle alignment becomes marginal in the core region of the channel. This orientational tendency is in accordance with earlier studies \citep{challabotla_shape_2015,zhao_rotation_2015}. It is noteworthy that particles exhibit a stronger preference and anisotropy in two-way coupled simulations. The two sets of disks collapse quite well concerning orientational statistics in one-way coupled simulations, while the modulated flows create notable differences.

\begin{figure}
	\flushright
	\begin{tabular}{ccc}	
		\begin{overpic}[width=3.9cm
			]{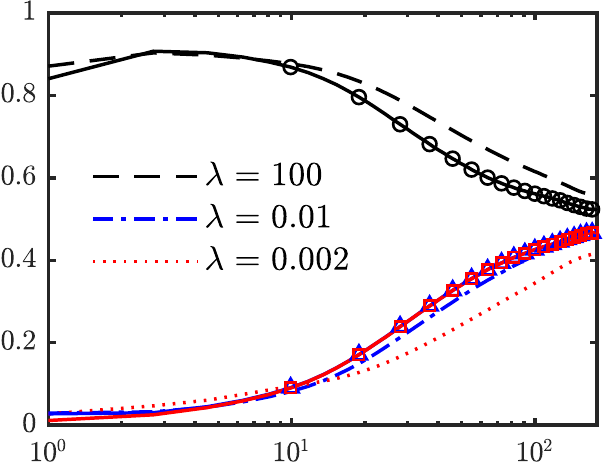}
			\put(-11,68){\textbf{(a)}} 
			\put(-13,25){ \rotatebox{90}{$<|n_i|>$}}			 
			\put(50,-10){\text {$z^+$}}
		\end{overpic}
		\hspace{0.2cm} 
		
		&
		\begin{overpic}[width=3.9cm
			]{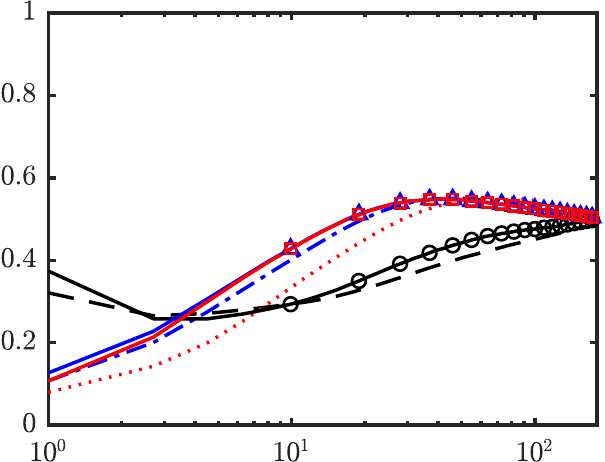}
			\put(-11,68){\textbf{(b)}} 
			\put(50,-10){\text {$z^+$}} 
		\end{overpic} 
		\hspace{0.2cm} 
		&
		\begin{overpic}[width=3.9cm
			]{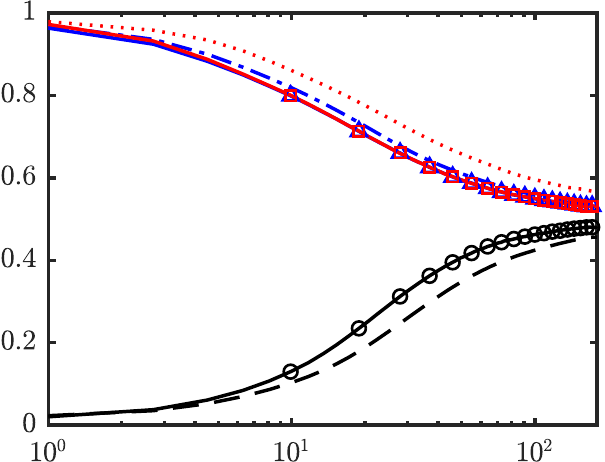}
			\put(-11,68){\textbf{(c)}} 
			\put(50,-10){\text {$z^+$}} 
		\end{overpic}   \\
	\end{tabular}
	\vspace{0.4cm}
	\caption{Mean absolute values of the direction cosines for spheroidal particles. (a) $\langle|n_x|\rangle$; (b)  $\langle|n_y|\rangle$; (c)  $\langle|n_z|\rangle$.  The solid lines with symbols represent particle statistics of one-way coupled simulations.}
	\label{fig6}
\end{figure}
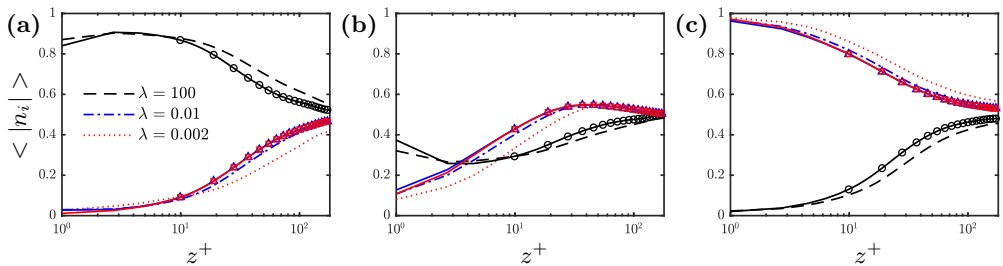

The rotational dynamics of inertialess spheroids are dominated by preferential particle orientation and fluid gradients (including mean shear and turbulent vorticity) \citep{yang2018Mean,zhao_mapping_2019}. As shown in Figure 7(a), rods rotate faster than disks with respect to mean spanwise angular velocity close to the wall, much slower than the fluid. The spin fluctuations of rods are larger than that of disks in the streamwise and spanwise directions, contrary to the tendency in the wall-normal direction. This spin anisotropy in the near-wall region is then related to the preferential particle orientation (see Figure 6). In consideration of the fluid vorticity field in Figure 5, it is not surprising to find the significant attenuation of $\langle \omega^+_y \rangle $ and rms$(\omega^{'+}_i)$ of spheroids in the particle-laden flows. Again, discrepancies between the rotational behavior of disks with $\lambda = 0.01$ and $\lambda = 0.002$ are enhanced due to particle-turbulence interactions. 

\begin{figure}
	\centering
	\begin{tabular}{ccc}	
		\begin{overpic}[width=6.5cm
			]{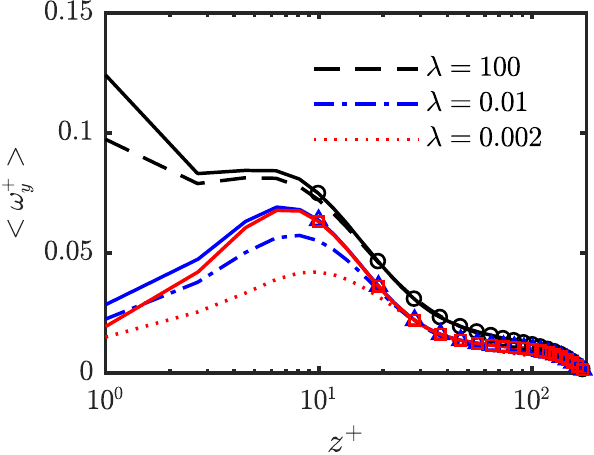}
			\put(0,70){\textbf{(a)}} 
		\end{overpic}  
		&
		\begin{overpic}[width=6.5cm
			]{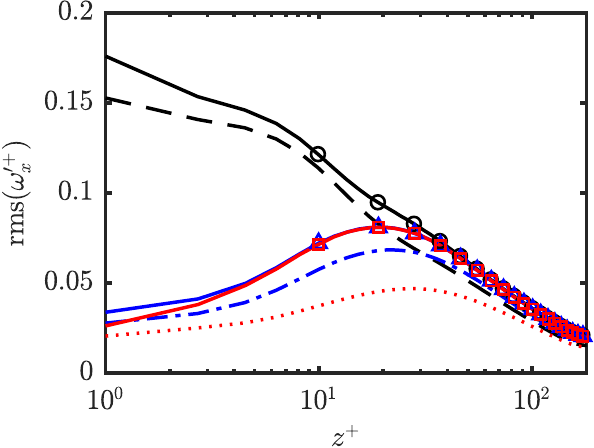}
			\put(0,70){\textbf{(b)}}
		\end{overpic} \\
		\begin{overpic}[width=6.5cm
			]{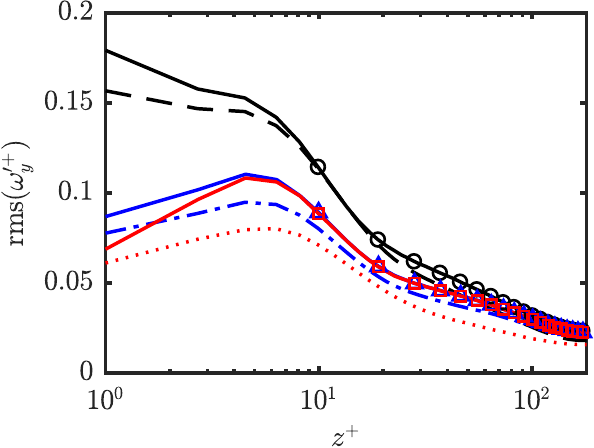}
			\put(0,70){\textbf{(c)}} 
		\end{overpic}  
		&
		\begin{overpic}[width=6.5cm
			]{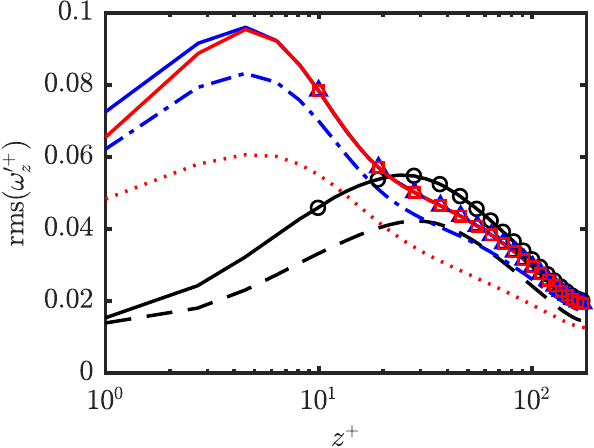}
			\put(0,70){\textbf{(d)}} 
		\end{overpic} \\
	\end{tabular}
	\caption{(a) Mean spanwise angular velocity; (b-d) Root-mean-square of the fluctuating angular velocity of particles in the streamwise, spanwise, and wall-normal direction, respectively. The solid lines with symbols represent particle statistics of one-way coupled simulations.}
	\label{fig7}			
\end{figure}

The variation of the mean particle stresses from the wall to the channel center is plotted in Figure 8. The components $\langle \tau^p_{xy} \rangle$ and $\langle \tau^p_{yz} \rangle$ vanish due to the symmetry of the channel flow. Different from the fluid viscous stresses, the appearance of mean particle normal stresses reveals that spheroids result in the non-Newtonian contribution to the carrier fluid. The mean particle stresses attain peaks in the buffer layer and vanish at the channel center. The normal stress $ \langle \tau^p_{xx} \rangle$ is slightly larger than other components for rods in the near-wall region, while the shear stress $\langle \tau^p_{xz} \rangle$ is the dominant one for disks. The present results show the same shape-dependence of particle stress with previous findings \citep{manhart_rheology_2003,wang_particle_2020,moosaie_rheology_2020} for spheroids with high asphericity in one-way coupled simulations. The maximum values of the normal stresses for the flattest disks are less than half of their counterparts for rods, while the opposite trend is observed for the shear stress. It seems that large shear stress $\tau^p_{xz}$ caused by disks is accompanied with pronounced turbulence modulations, e.g. the fluid vorticity and turbulence intensities, but the degree of drag reduction is less than expected. The distribution of mean particle stress for disks is qualitatively different from that for rods. Further details about the effect of particle stress on the fluid phase will be explored in the following section.

\begin{figure}
	\flushright
	\begin{tabular}{ccc}	
		\begin{overpic}[width=3.9cm
			]{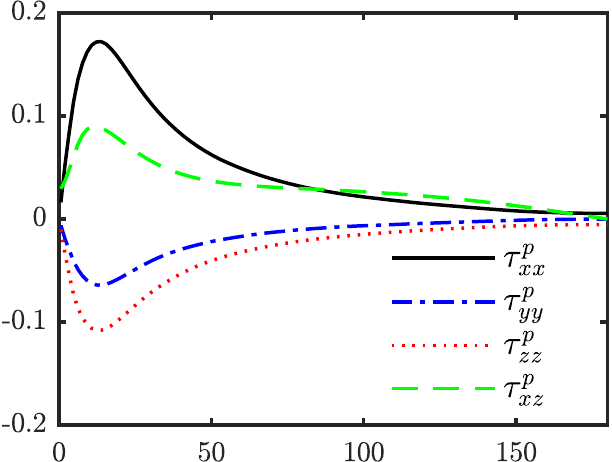}
			\put(-11,68){\textbf{(a)}} 
			\put(-13,25){ \rotatebox{90}{$<\tau^{p+}_{ij}>$}}			 
			\put(50,-10){\text {$z^+$}}
		\end{overpic}
		\hspace{0.2cm} 
		
		&
		\begin{overpic}[width=3.9cm
			]{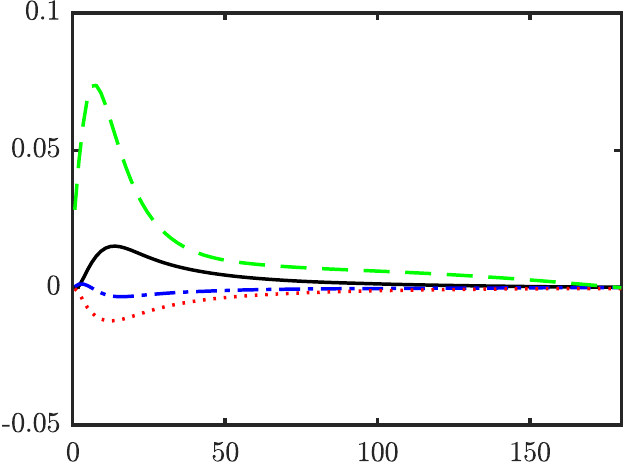}
			\put(-11,68){\textbf{(b)}} 
			\put(50,-10){\text {$z^+$}} 
		\end{overpic} 
		\hspace{0.2cm} 
		&
		\begin{overpic}[width=3.9cm
			]{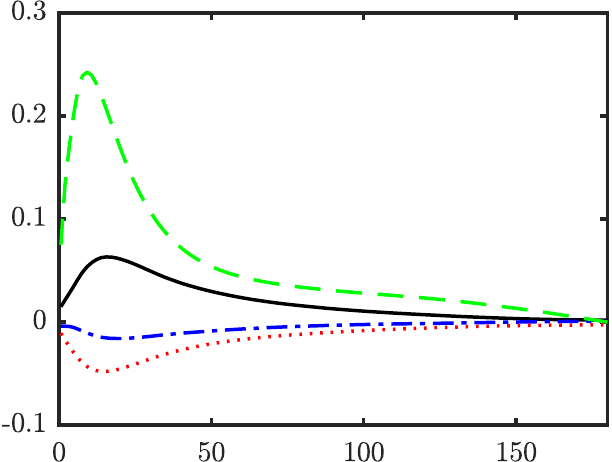}
			\put(-11,68){\textbf{(c)}} 
			\put(50,-10){\text {$z^+$}} 
		\end{overpic}   \\
	\end{tabular}
	\vspace{0.4cm}
		\caption{Mean particle stresses in two-way coupled simulations for spheroids with (a) $\lambda =100$, (b) $\lambda =0.01$, and (c) $\lambda = 0.002$.}
		\label{fig8}
\end{figure}
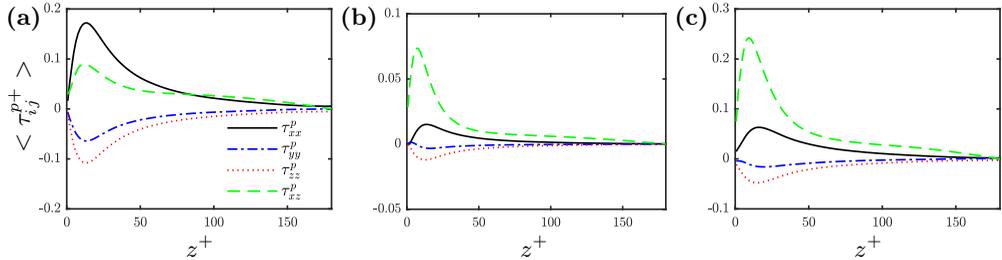

\subsection{Drag reduction mechanism}

In Section 3.1 and 3.2, we have presented the fluid statistics and the particle statistics in the suspensions of spheroids with high asphericity, compared with results in one-way coupled simulations. However, two questions remain unanswered. Firstly, how do inertialess spheroids modulate the turbulence field, especially coherent structures? Secondly, why rods are more effective than disks in drag reduction? Motivated by those questions, in this section, we will explore the mechanisms for drag reduction by inertialess spheroids in turbulent channel flow.

The near-wall coherent structures formed in the self-sustaining process are responsible for high skin-friction drag in wall-bounded turbulence \citep{kravchenko_relation_1993,hamilton_regeneration_1995}. Modulating the turbulence structures, especially the quasistreamwise vortices, is an appropriate way to induce turbulent drag reduction \citep{kim_physics_2011}.
The effect of inertialess spheroids on instantaneous flow fields is examined at first. Figure 9 illustrates the vortex structures in the lower half channel visualized by the isosurfaces of the $\lambda_2$-criterion \citep{Jeong1995On} and colored by the fluctuating streamwise velocity. In the near-wall region, the dominant structures are the quasistreamwise vortices, slightly tilted away from the wall. Compared with the unladen flow case in Figure 9(a), the addition of spheroids results in fewer vortices with larger sizes. Disks with $\lambda = 0.002$ generate the most pronounced modulation, followed by rods with $\lambda = 100$, and finally disks with $\lambda = 0.01$. These phenomena are consistent with the suppression of the fluctuating fluid vorticity and partly explain why the turbulence intensities in the spanwise and wall-normal directions are attenuated. \cite{kim_physics_2011} pointed out that the attenuation of the quasistreamwise vortices is a common feature of drag-reduced  wall-bounded flows. 

 \begin{figure}
	\centering
	\begin{tabular}{ccc}	
		\begin{overpic}[width=6.2cm
			]{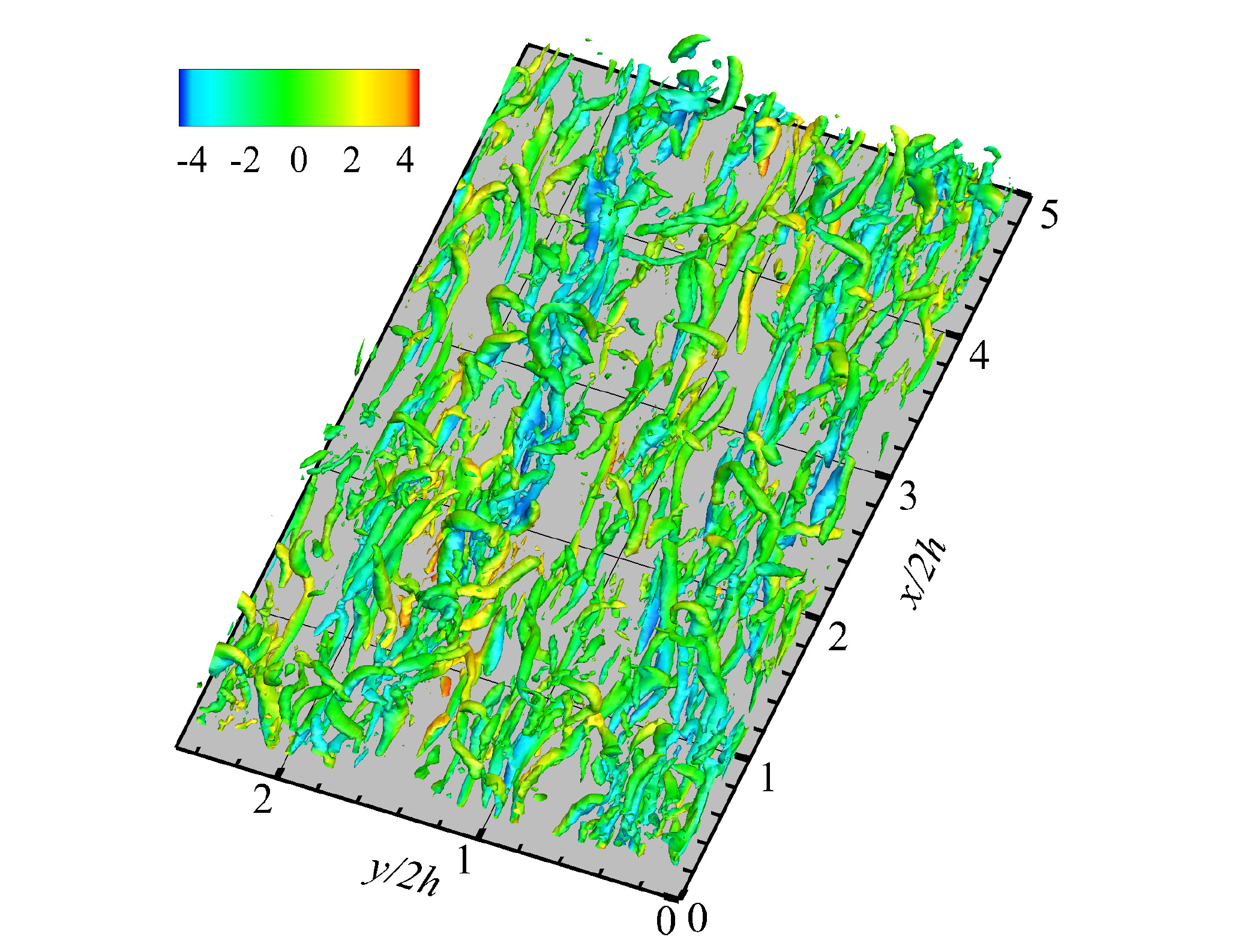}
			\put(0,55){\textbf{(a)}} 
			
			\put(5,67){\textbf{$u'^+$}}
		\end{overpic}  
		&
		\begin{overpic}[width=6.2cm
			]{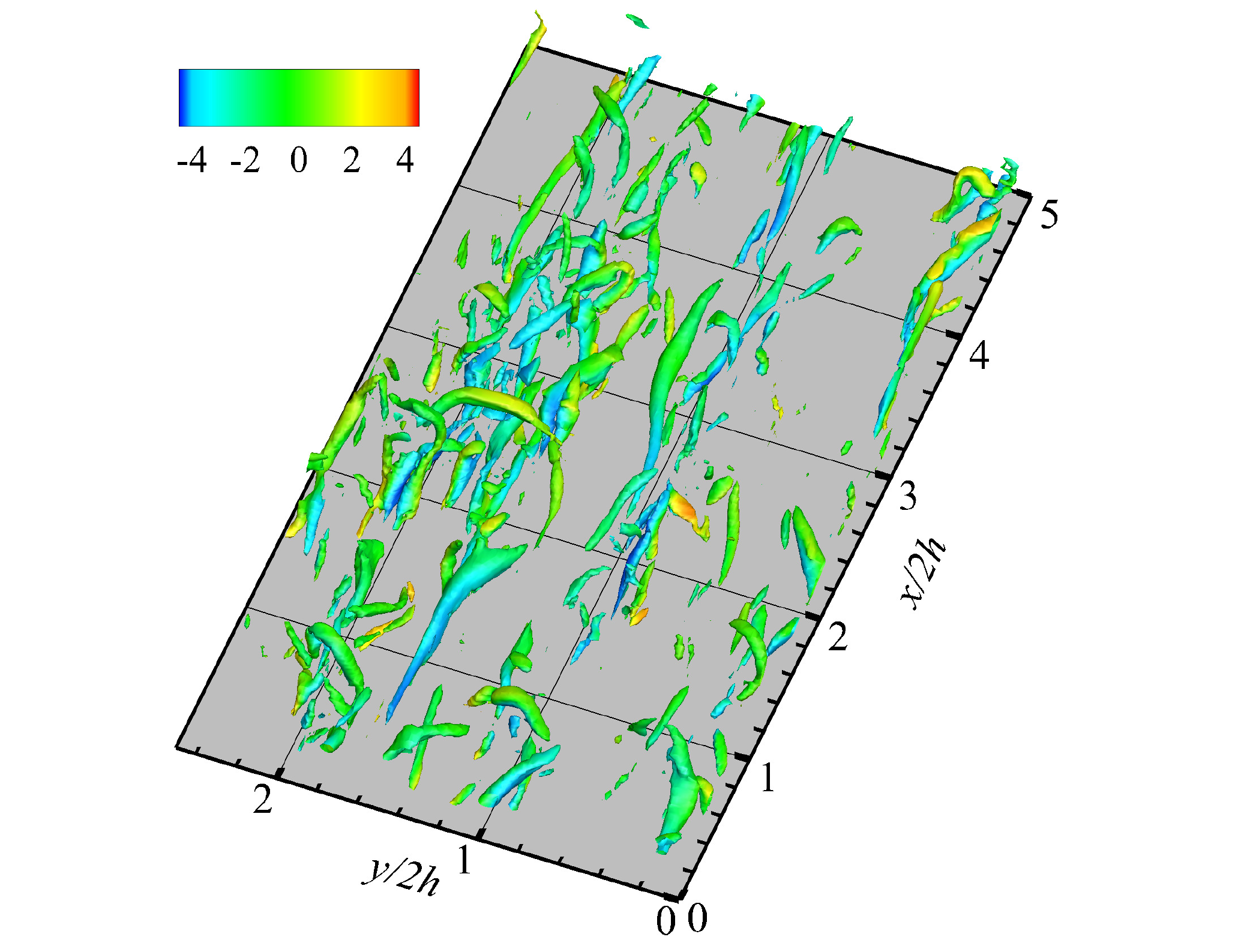}
			\put(0,55){\textbf{(b)}}
			\put(5,67){\textbf{$u'^+$}}
		\end{overpic} \\
		\begin{overpic}[width=6.2cm
			]{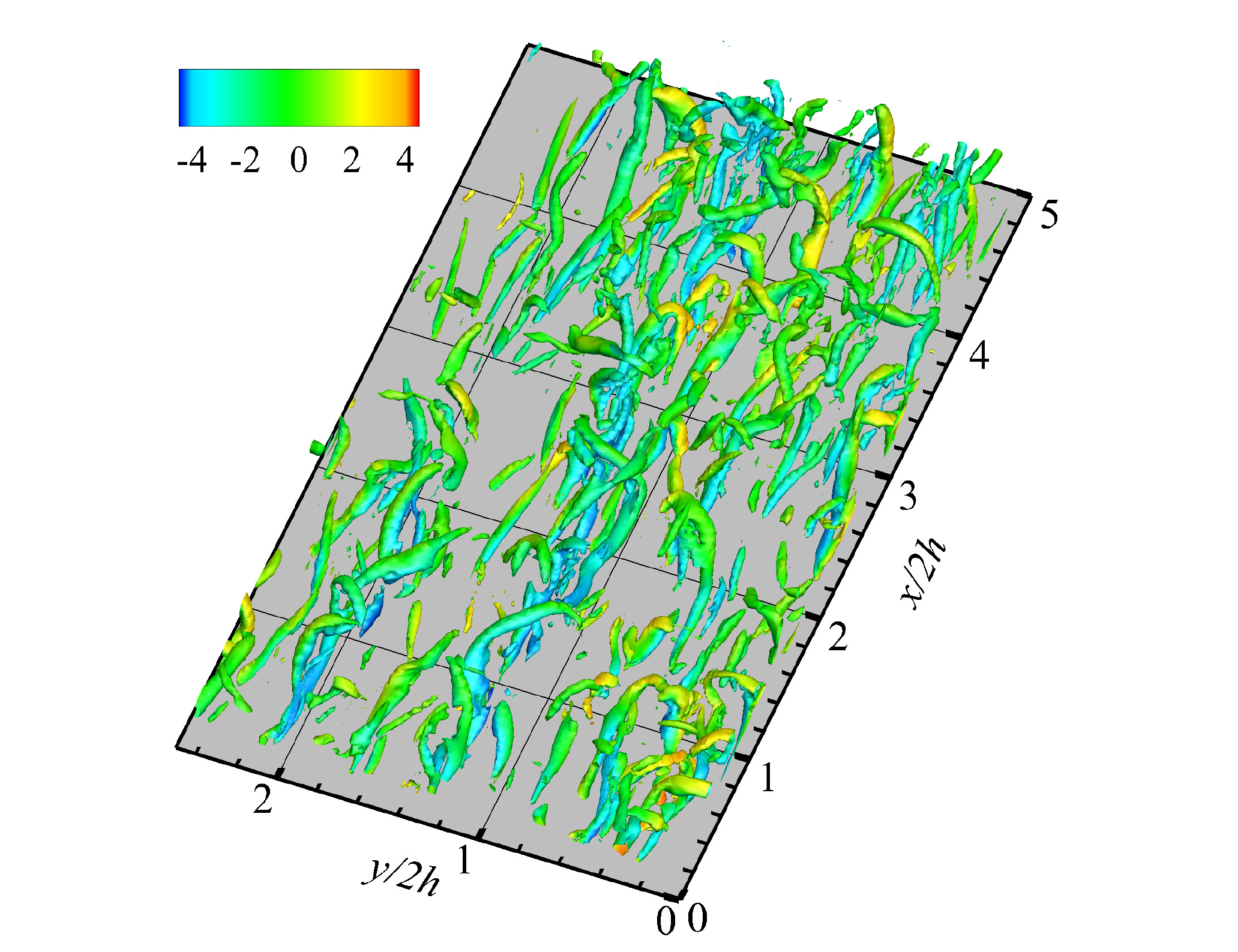}
			\put(0,55){\textbf{(c)}} 
			
			\put(5,67){\textbf{$u'^+$}}
			
		\end{overpic}  
		&
		\begin{overpic}[width=6.2cm
			]{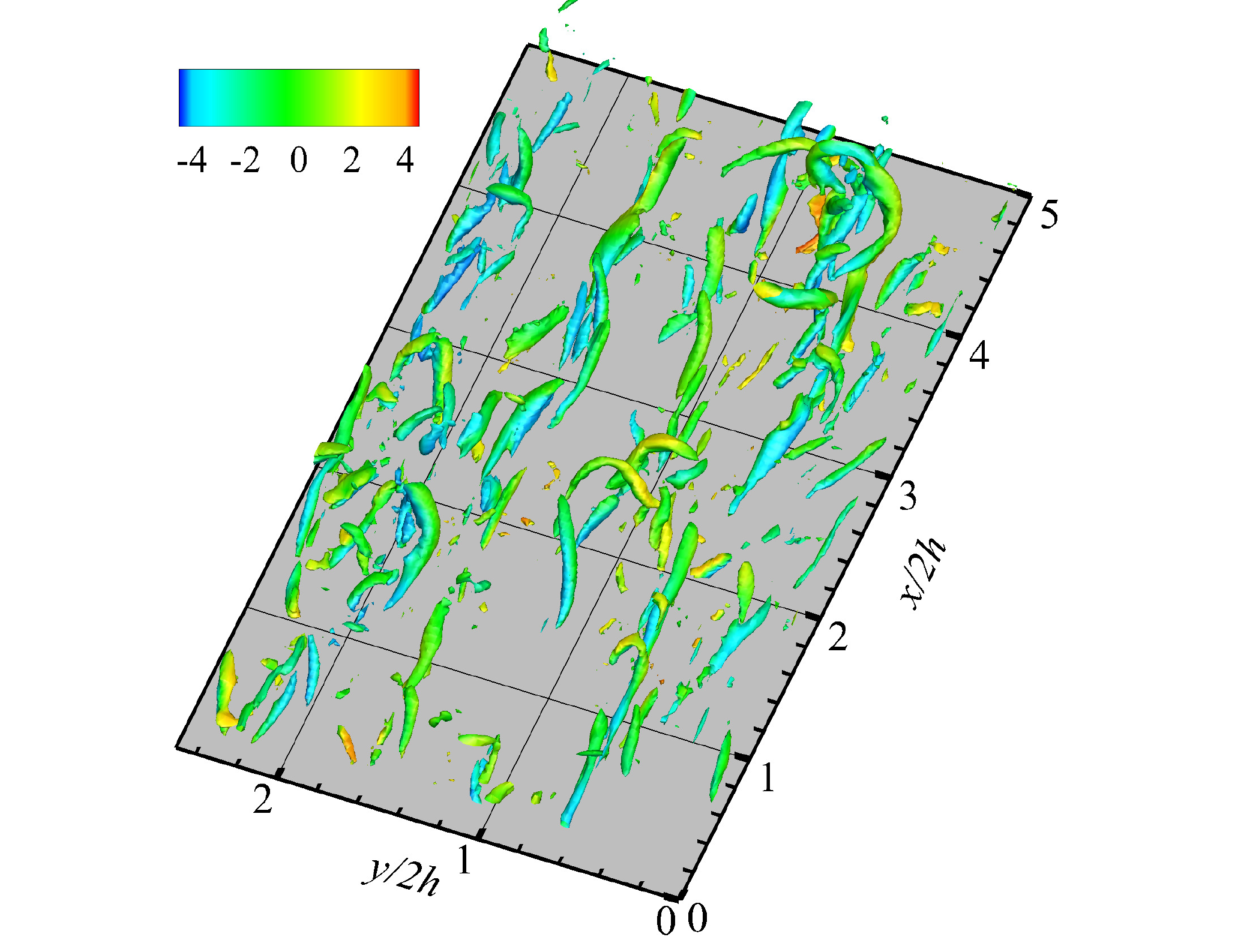}
			\put(0,55){\textbf{(d)}} 
			
			\put(5,67){\textbf{$u'^+$}}
		\end{overpic} \\
	\end{tabular}
	\caption{Instantaneous isosurfaces of $\lambda_2^+ = -0.007$ colored by streamwise velocity $u'^+$ in the lower half channel for (a) unladen flow and particle-laden flows with (b) $\lambda =100$; (c) $\lambda =0.01$; (d) $\lambda =0.002$.}
	\label{fig9}			
\end{figure}
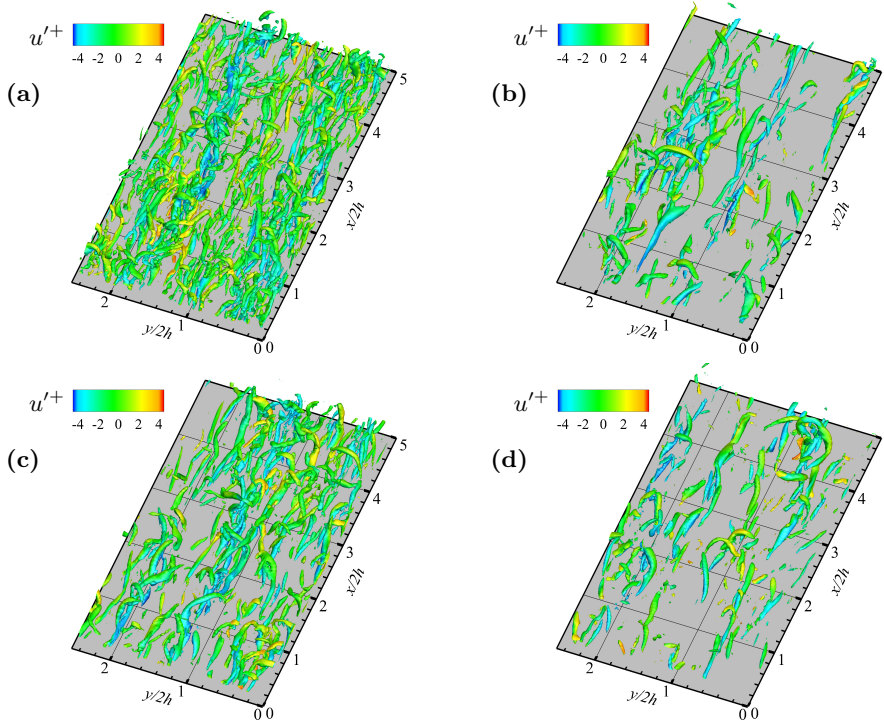

Produced by the quasistreamwise vortices, the near-wall streaky structures should respond to the addition of spheroids. From Figure 10, we observe that streaks become more regular and the spanwise spacing is increased in particle suspensions. The reduction of waviness and small scales means that the low-speed streaks are more stable, hence the generation of the quasistreamwise vortices is suppressed \citep{SCHOPPA2002Coherent}. 
The mean streak spacing is further investigated through velocity autocorrelation in the spanwise direction. The separation of minimum correlation corresponds to half of the mean streak spacing \citep{kim_turbulence_1987}. In the near-wall region $z^+<40$, the spacing for particle suspensions is larger than that in unladen flow, but the differences gradually diminish as away from the wall (see Figure 11). Since the near-wall streak spacing is independent of Reynolds number \citep{smits_highreynolds_2011}, Figure 11 also demonstrates that the flow laden with spheroids behaves differently to the Newtonian fluid.

   	\begin{figure}
	\raggedleft
	\begin{tabular}{ccc}	
		\begin{overpic}[width=6.2cm
			]{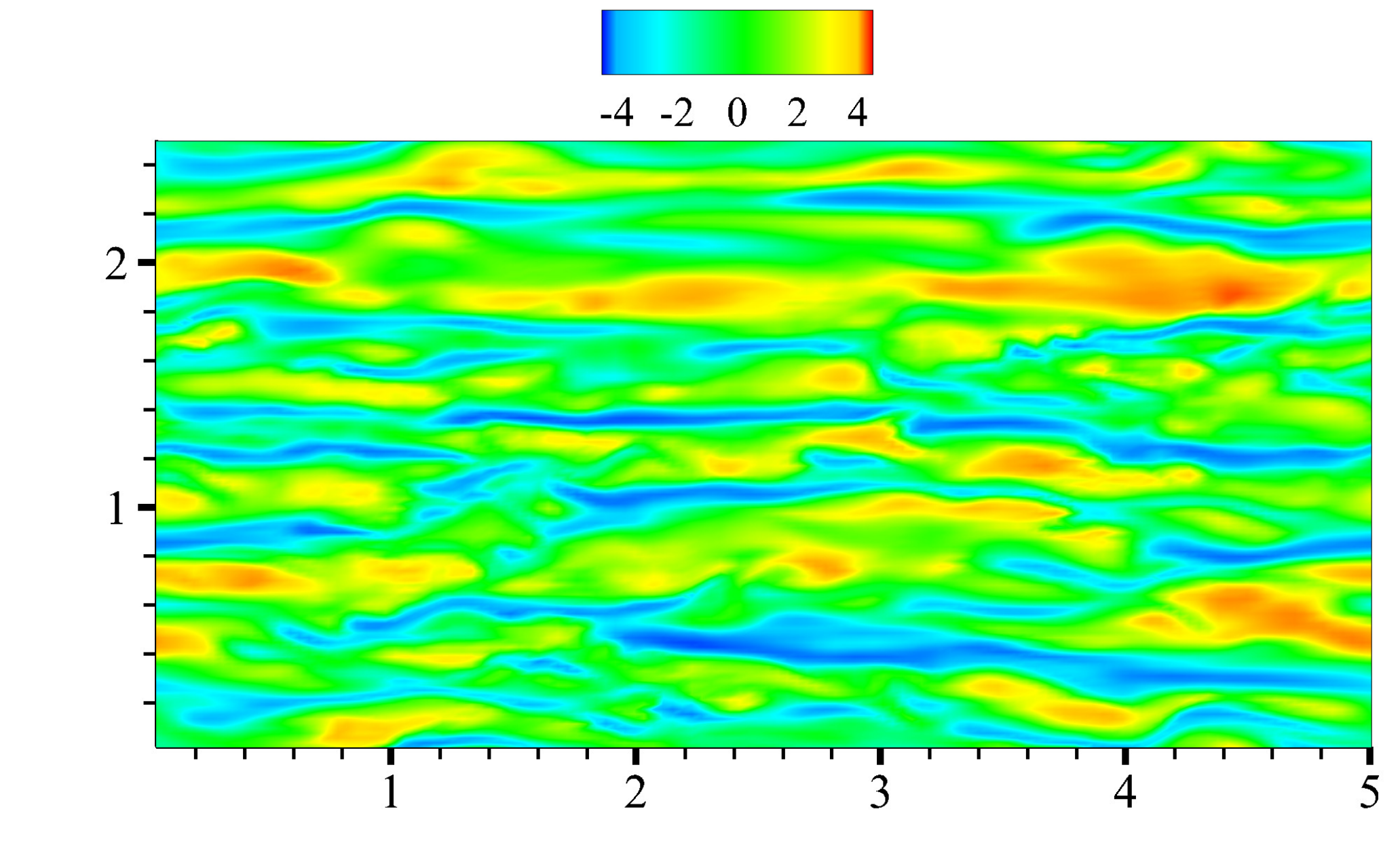}
			\put(0,55){\textbf{(a)}} 
			\put(-5,30){\textbf{$y/2h$}}
			\put(32,57){\textbf{$u'^+$}}
		\end{overpic}  
		&
		\begin{overpic}[width=6.2cm
			]{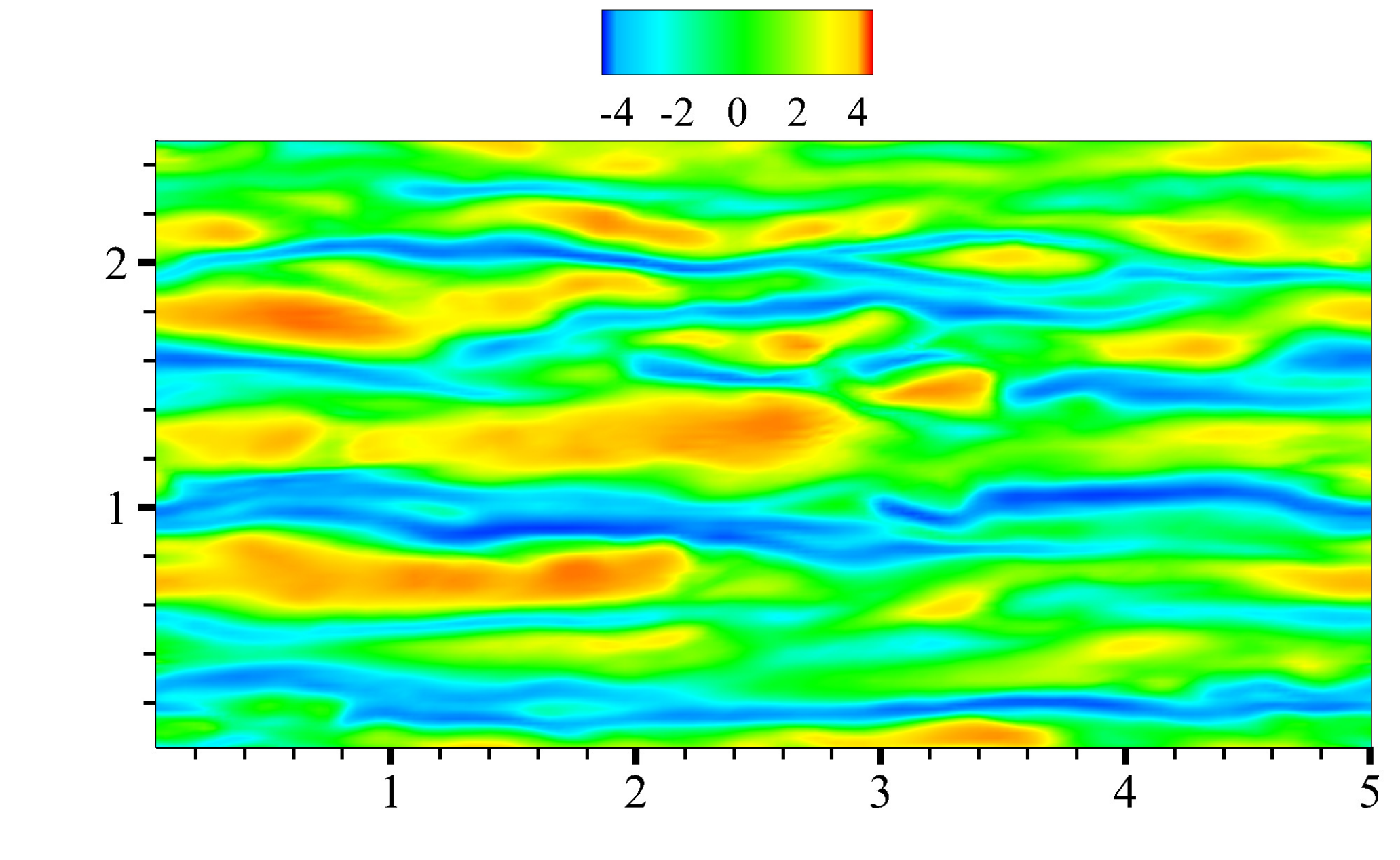}
			\put(0,55){\textbf{(b)}} 
			\put(32,57){\textbf{$u'^+$}}
		\end{overpic} \\
		\begin{overpic}[width=6.2cm
			]{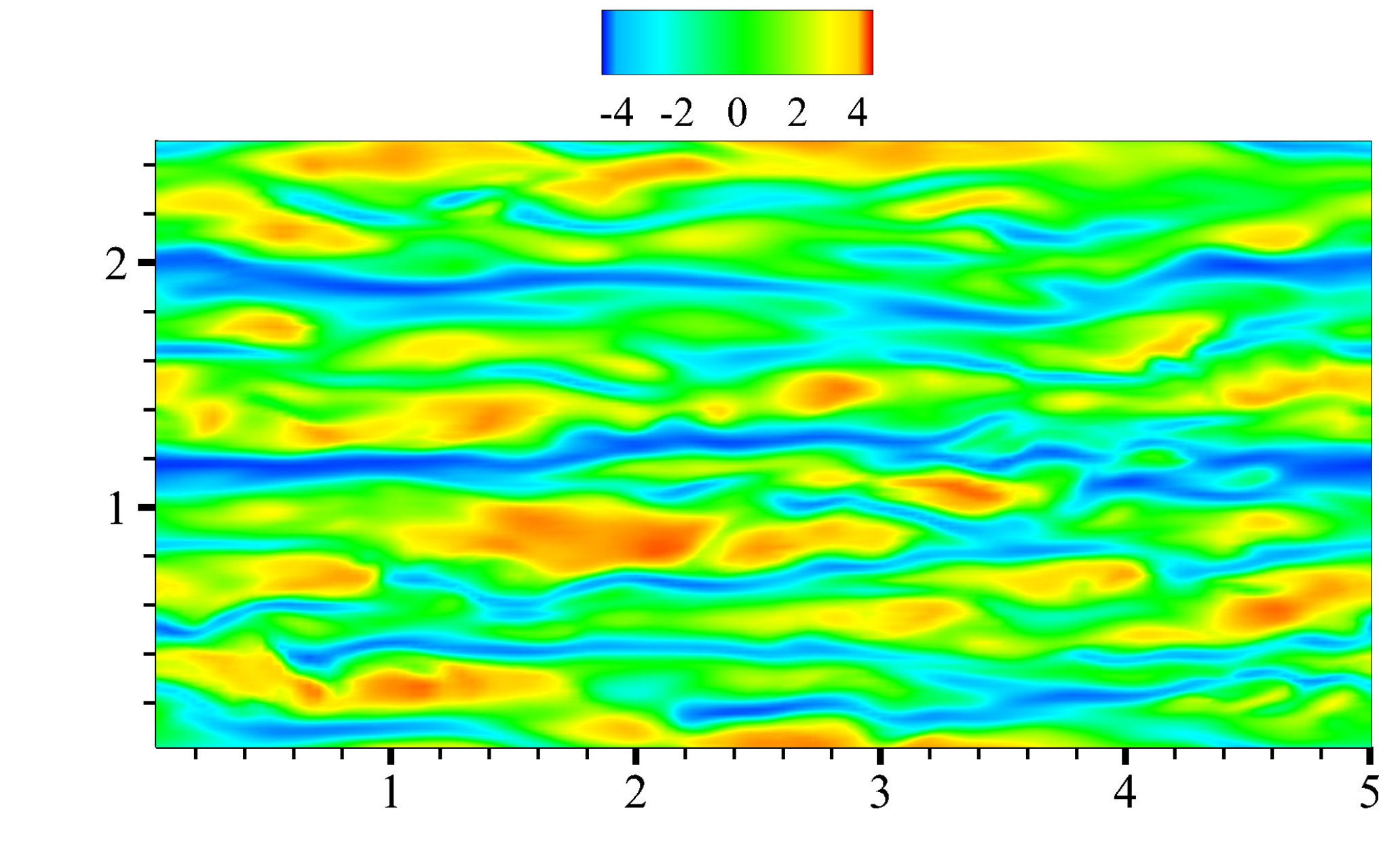}
			\put(0,55){\textbf{(c)}} 
			\put(-5,30){\textbf{$y/2h$}}
			\put(50,-2){\textbf{$x/2h$}} 
			\put(32,57){\textbf{$u'^+$}}
			
		\end{overpic}  
		&
		\begin{overpic}[width=6.2cm
			]{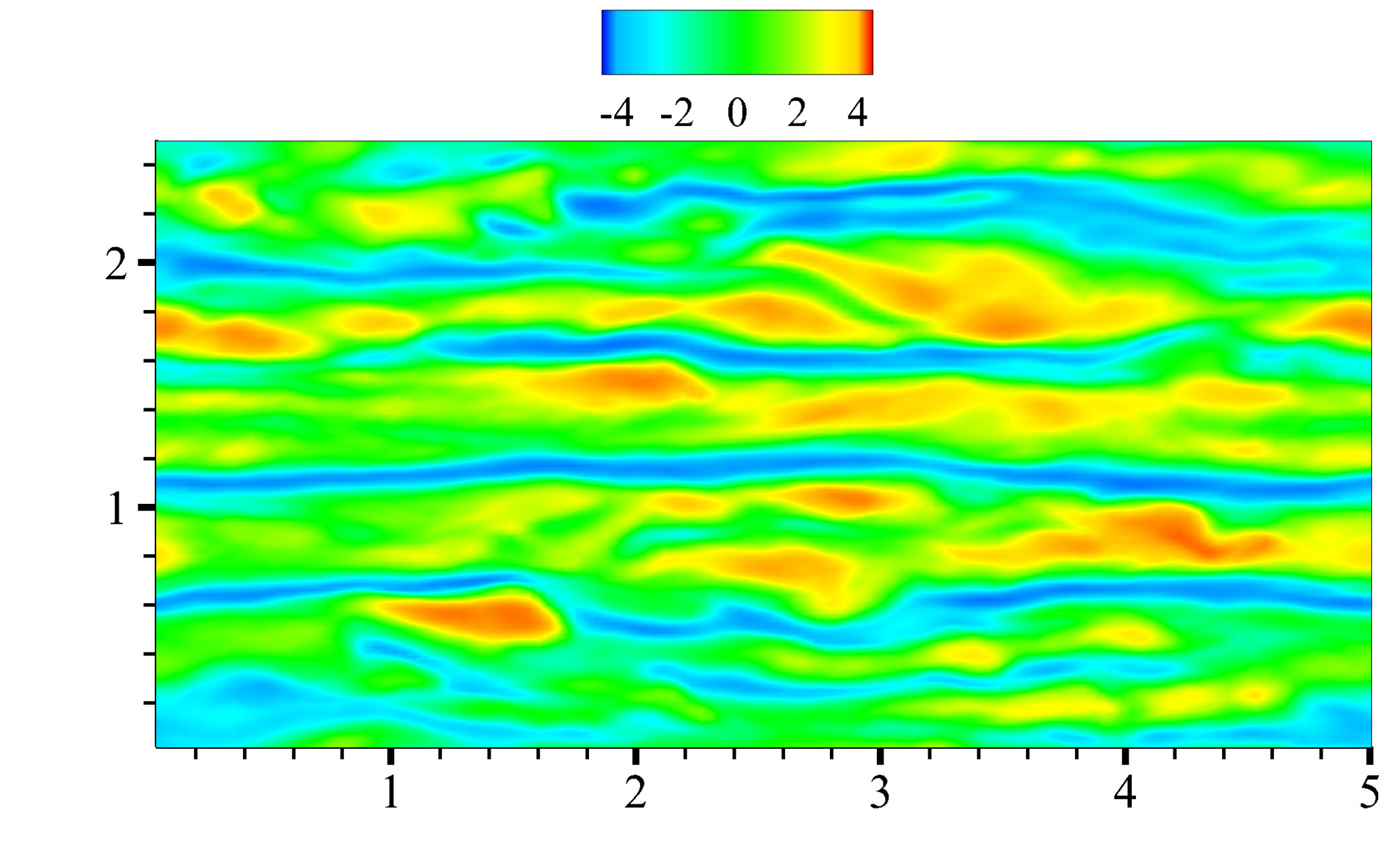}
			\put(0,55){\textbf{(d)}} 
			\put(50,-2){\textbf{$x/2h$}} 
			\put(32,57){\textbf{$u'^+$}}
		\end{overpic} \\
	\end{tabular}
	\caption{Instantaneous contours of streamwise velocity fluctuations at $z^+ = 15$ for (a) unladen flow and particle-laden flows with (b) $\lambda =100$; (c) $\lambda =0.01$; (d) $\lambda =0.002$.}
	\label{fig10}			
\end{figure}

\begin{figure}
	\centering
	\begin{overpic}[width=6.5cm
		]{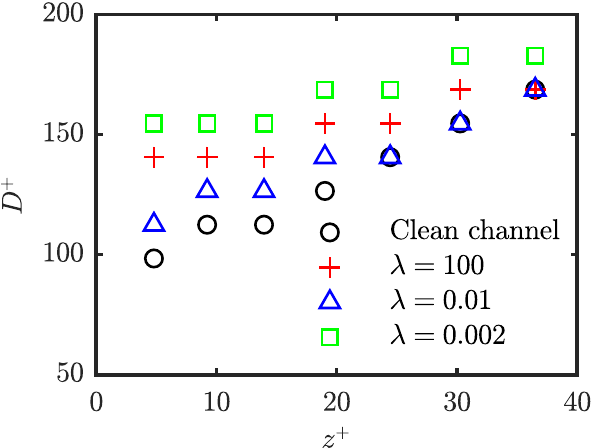}
	\end{overpic} 
	\caption{Variation of mean spanwise streak spacing in near-wall region.}
	\label{fig11}
\end{figure}

The passage of the quasistreamwise vortices generates the bursting events, accounting for the production of Reynolds shear stress. As a consequence, the weakened vortex structures leads to the reduction of Reynolds shear stress. This provides the possibility of turbulent drag reduction. Therefore, it is of significance to figure out how inertialess spheroids modulate vortices, the quasistreamwise vortices in particular. Here, the modulation is examined by the work done by particles on the fluid. According to Appendix A, the transport equation of the turbulent kinetic energy $k$ in particle-laden flow can be written as

\begin{equation}
\begin{split}
\frac{\partial{k}}{\partial{t}}+\langle u_i\rangle \frac{\partial{k}}{\partial{x_i}} 
&=-\langle u'_iu'_j \rangle \frac{\partial{\langle u_i\rangle }}{\partial{x_j}}
-\frac{\partial}{\partial{x_j}}\left(\frac{\langle p'u'_j\rangle}{\rho}+\frac{\langle u'_i u'_i u'_j\rangle}{2} -\nu\frac{\partial{k}}{\partial{x_j}}\right)\\
&\quad-\nu\left\langle\frac{\partial{u'_i}}{\partial{x_j}}\frac{\partial{u'_i}}{\partial{x_j}}\right\rangle
+\frac 1\rho\langle f'_iu'_i\rangle.
\end{split}
\label{turbulent kinetic energy balance equation}
\end{equation}
Here, $f_i = {\partial{\tau^{p}_{ij}}}/{\partial{x_j}}$ is particle body force \citep{paschkewitz_numerical_2004}. The work done by particles on the fluid per mass and per time in the $\gamma (\gamma=x, y, z)$ direction is then defined as $W_{\gamma}=\langle f'_{\gamma}u'_{\gamma} \rangle$.
Note that the usual summation convention is not adopted for the repeated Greek indices. The work done by particles on the fluid has also been analyzed in earlier studies \citep{paschkewitz_numerical_2004,paschkewitz_dynamic_2005,zhao_interphasial_2013,lee_modification_2015,pan_kinetic_2020}. We focus on the modulation of the quasistreamwise vortices, which mainly consist of velocities in the spanwise and wall-normal directions, and the work $W_{y+z}=W_y+W_z$ is taken into account. The negative (positive) $W_{y+z}$ indicates that particles weaken (strengthen) the quasistreamwise vortices. The sign of $W_{y+z}$ is determined by the angle between two vectors of  $\mathbf{v'}+\mathbf{w}'$ and $\mathbf{f'_y}+\mathbf{f'_z}$.

Figure 12 plots the probability density functions (PDFs) of $W^+_{y+z}$ conditionally sampled in the region $20<z^+<60$ where the self-sustaining process happens. The vortices are extracted from background fluctuations with the condition $\lambda_2<-\lambda_{2,\mathrm{rms}}$. We observe that the probability of negative work is higher than that of positive work, regardless of the particle shape. This observation reveals that the work done by spheroids contribute to the attenuation of velocity fluctuations in the y and z directions in the vortex regions. However, rods and the flattest disks, with the less sharp profiles of PDFs, are more likely to produce large work than disks with $\lambda = 0.01$. 

\begin{figure}
	\centering
	\begin{overpic}[width=6.5cm
		]{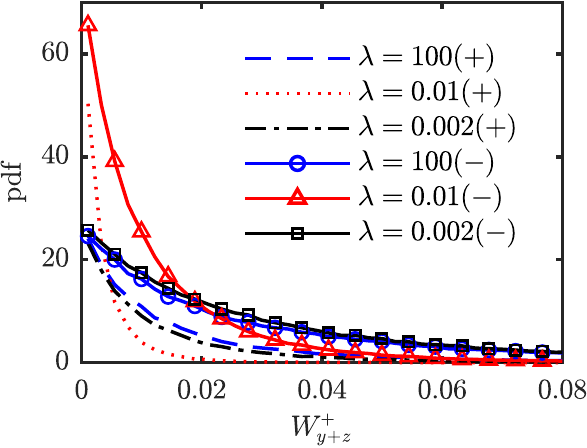}
	\end{overpic} 
	\caption{Probability density functions of the work $W^+_{y+z}$ done by particles on the fluid conditionally sampled with $\lambda_2<-\lambda_{2,\mathrm{rms}}$ in $20<z^+<60$. The solid lines with symbols represent negative values.}
	\label{fig12}
\end{figure}

In order to pronounce the effect on the quasistreamwise vortices, the left column panels of Figure 13 show the instantaneous three-dimensional isosurfaces of $\lambda_2$ in half channel and slices of $W^+_{y+z}$ in the y-z and x-z planes. It is clear that spheroids tend to produce negative work on the vortex structures. From slices of $W^+_{y+z}$, the large negative work is associated with the regions near the quasistreamwise vortices. The magnitude of $W^+_{y+z}$ done by disks with $\lambda = 0.01$ is mostly smaller than that by other spheroids, consistent with PDFs in Figure 12. The instantaneous results manifest that $W_{y+z}$ done by spheroids weakens the quasistreamwise vortices.

 \begin{figure}
	\centering
	\begin{tabular}{ccc}	
		\begin{overpic}[width=6.2cm
			]{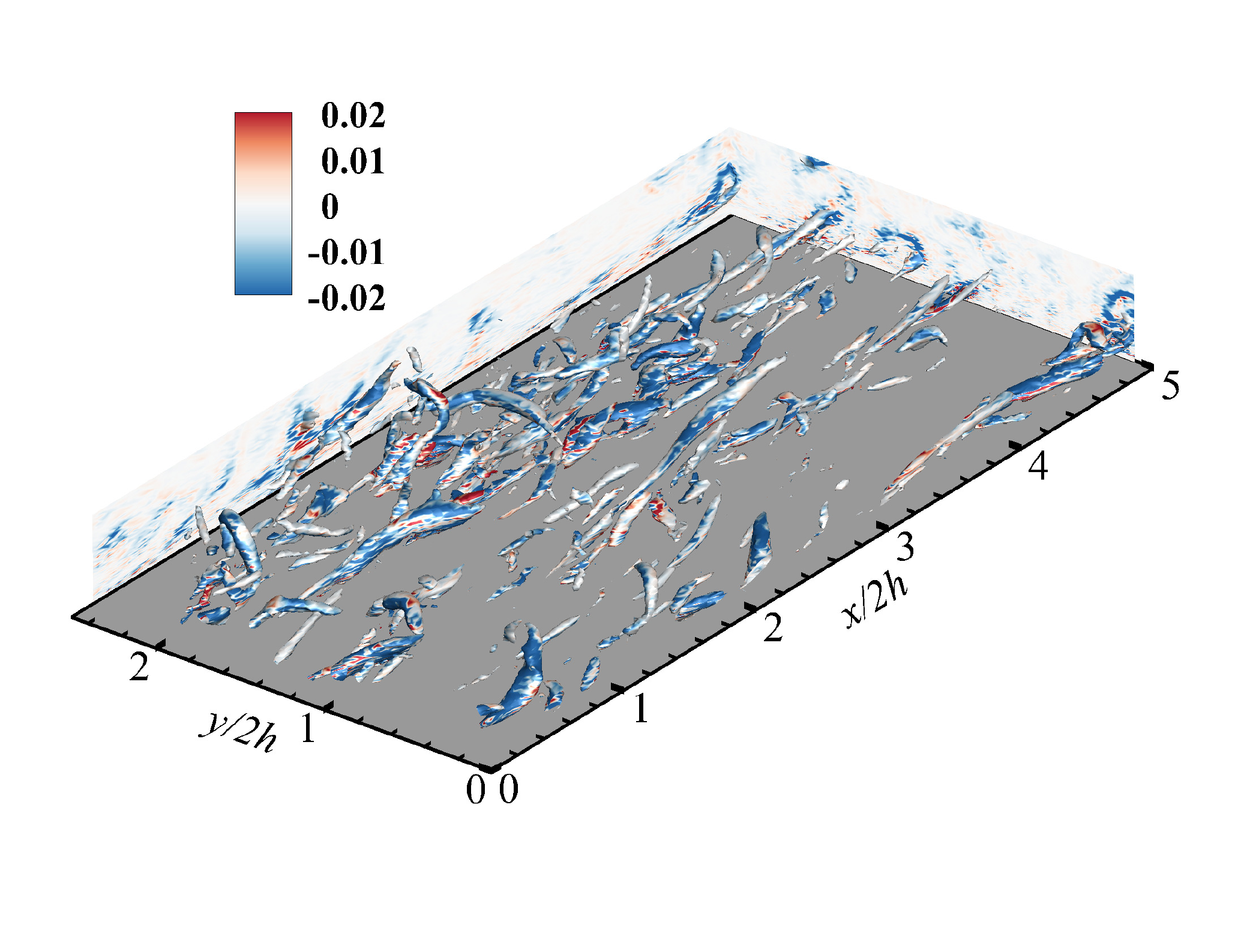}
			\put(0,70){\textbf{(a)}} 
			
			\put(2,57){\textbf{$W^+_{y+z}$}}
		\end{overpic}  
		&  
		\begin{overpic}[width=6.2cm
			]{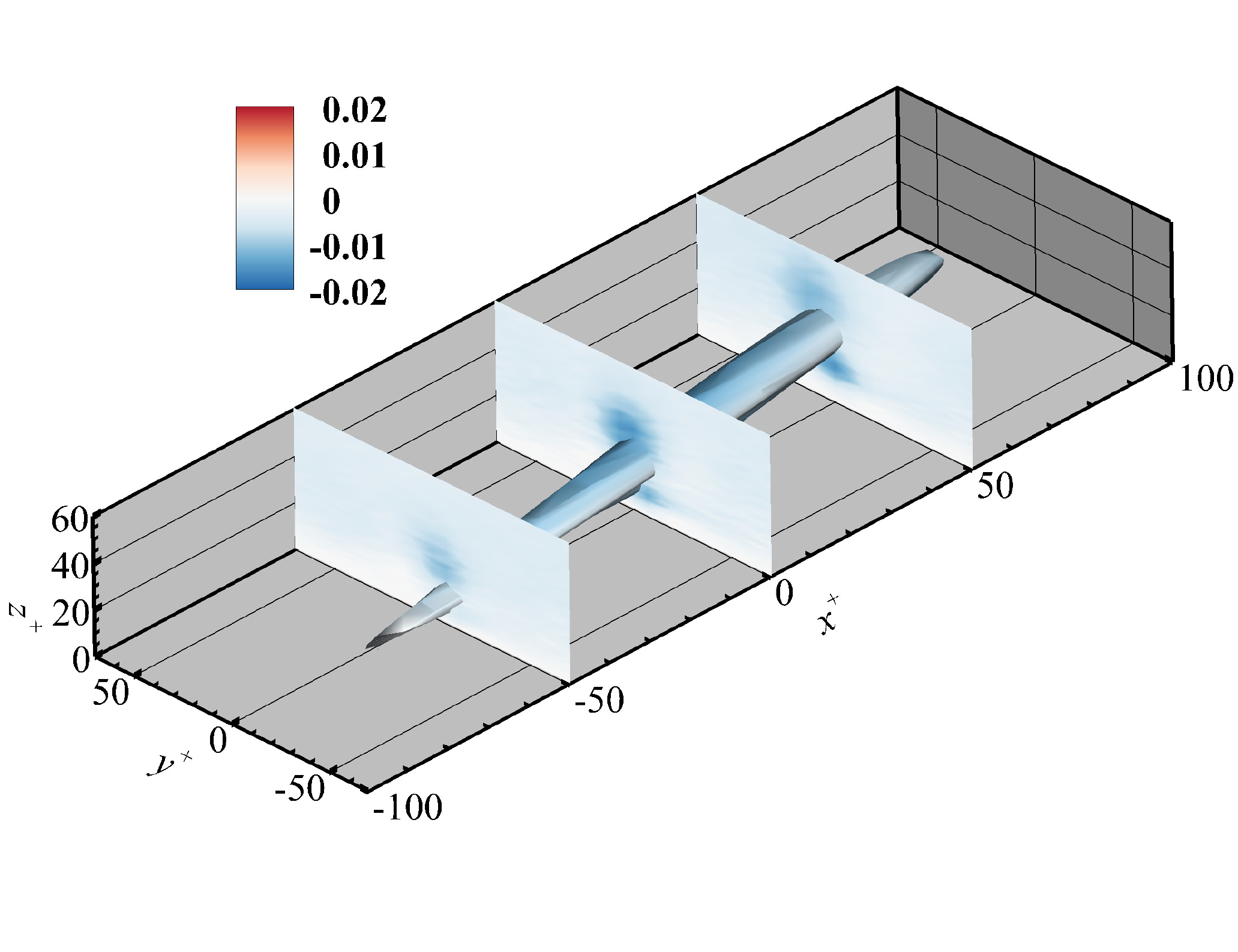}
			\put(0,70){\textbf{(b)}} 
			
			\put(5,57){\textbf{$W^+_{y+z}$}}
		\end{overpic}  \\
		
		\begin{overpic}[width=6.2cm
			]{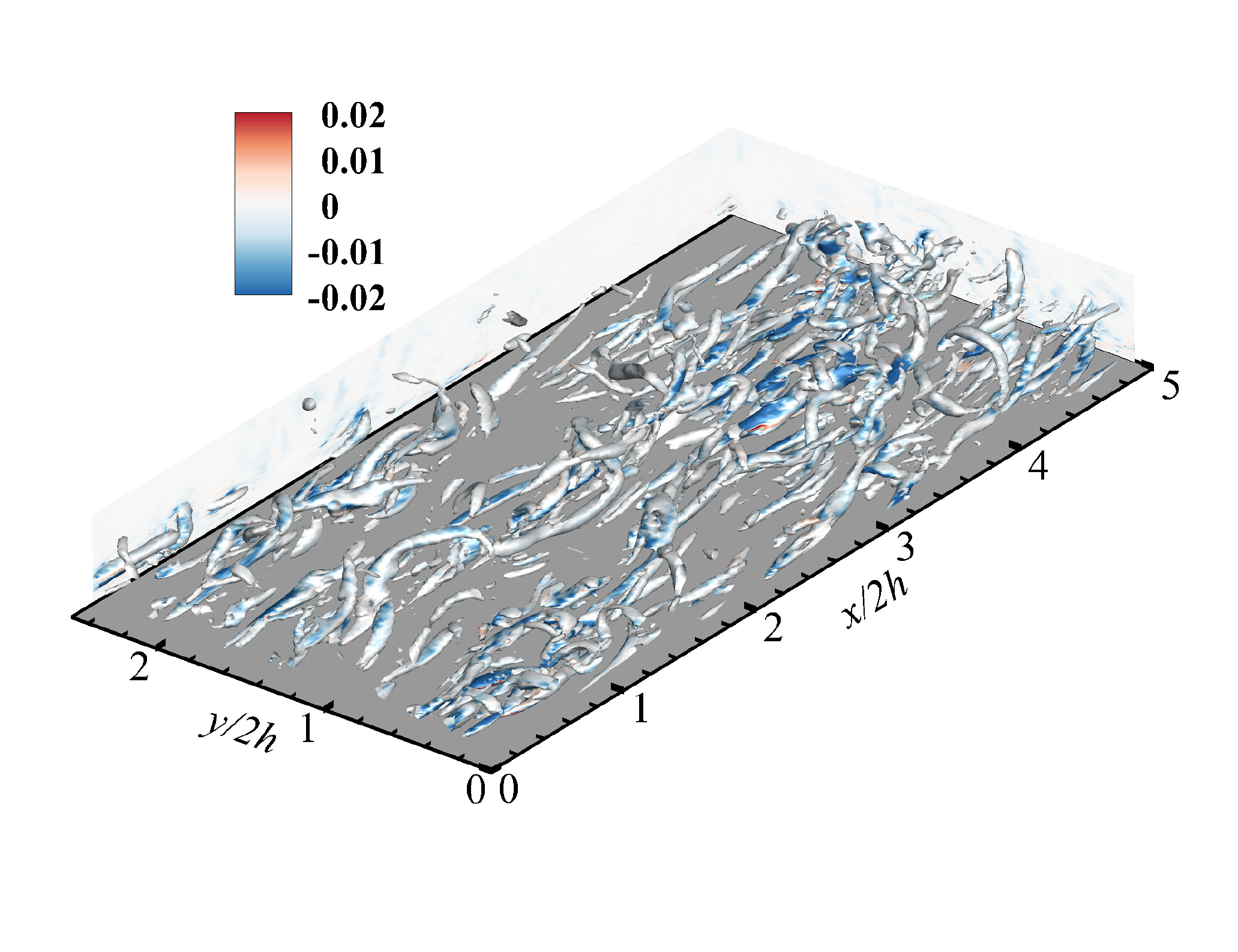}
			\put(0,70){\textbf{(c)}} 
			\put(2,57){\textbf{$W^+_{y+z}$}}
		\end{overpic} 
		&
		\begin{overpic}[width=6.2cm
			]{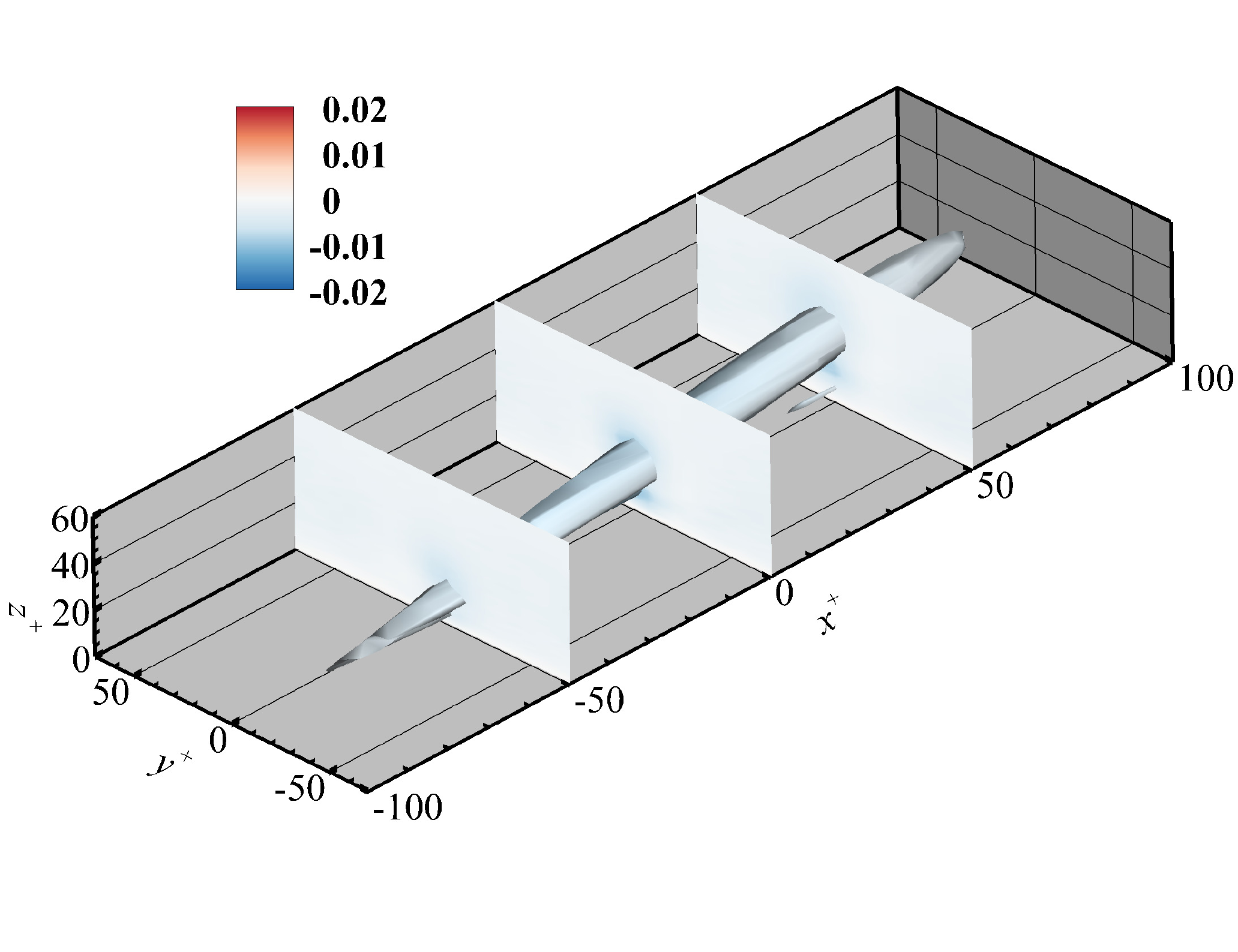}
			\put(0,70){\textbf{(d)}} 
			\put(5,57){\textbf{$W^+_{y+z}$}}
		\end{overpic} \\
		\begin{overpic}[width=6.2cm
			]{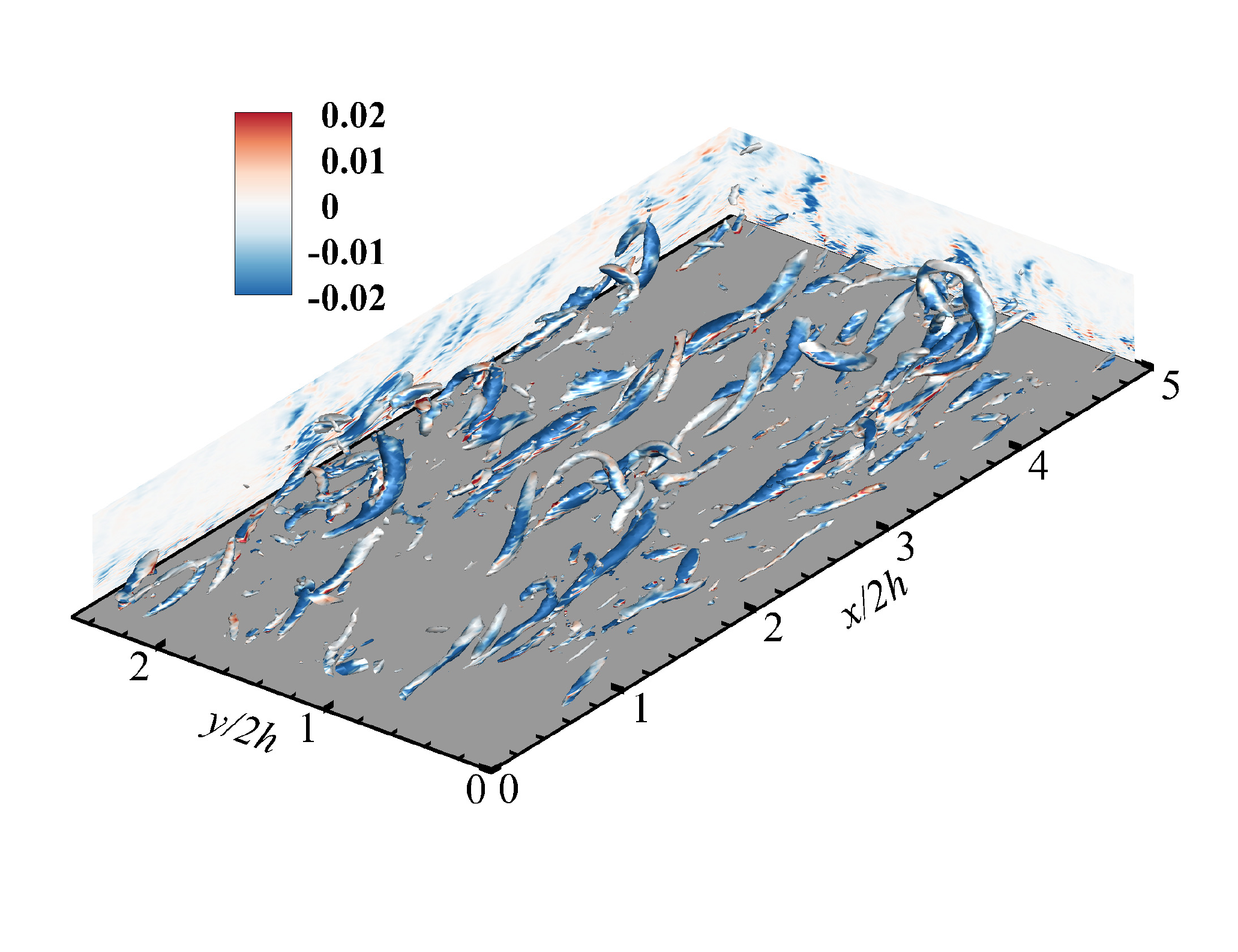}
			\put(0,70){\textbf{(e)}} 
			
			\put(2,57){\textbf{$W^+_{y+z}$}}
		\end{overpic} 
		&
		\begin{overpic}[width=6.2cm
			]{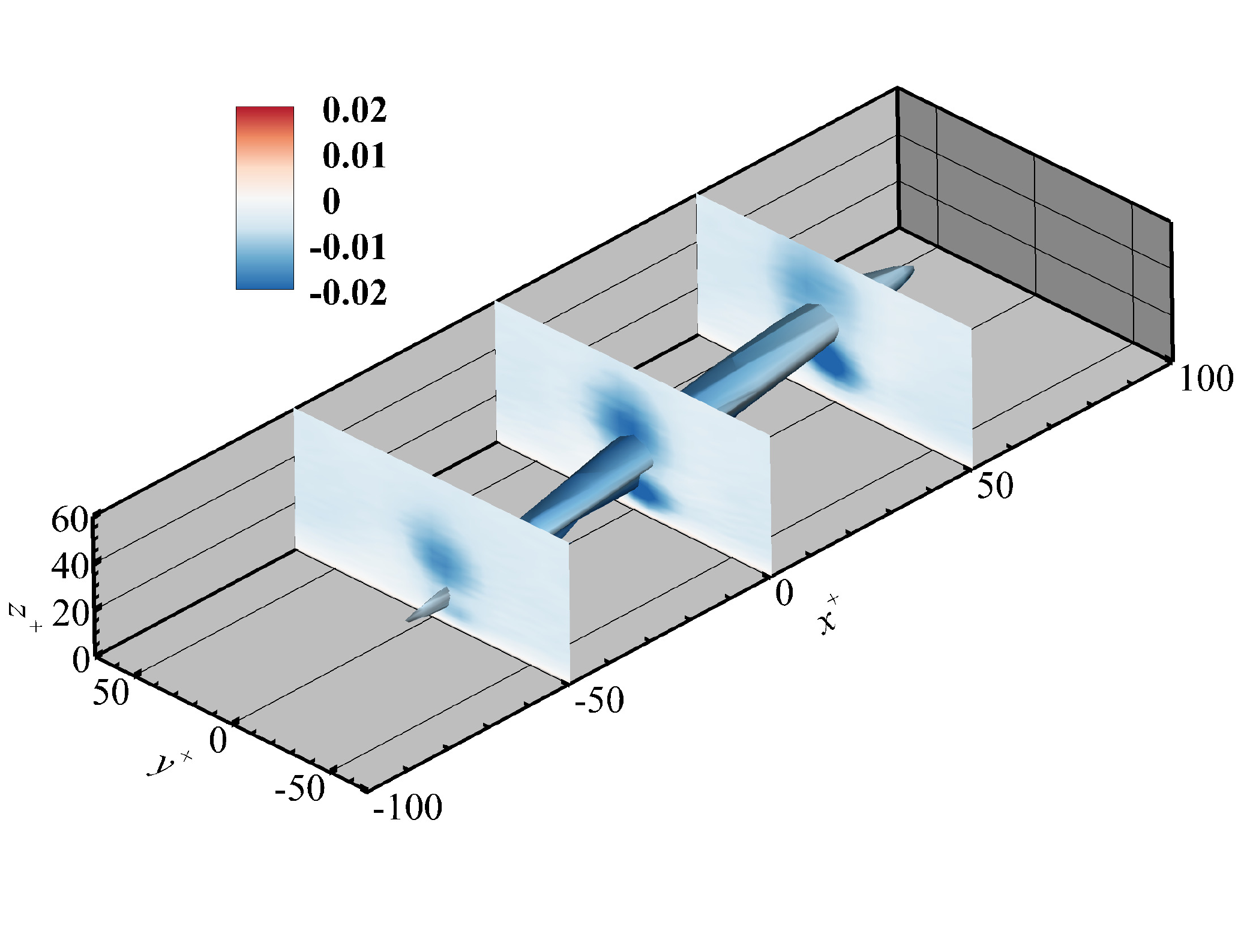}
			\put(0,70){\textbf{(f)}} 
			
			\put(5,57){\textbf{$W^+_{y+z}$}}

		\end{overpic}
		\\
	\end{tabular}
	\caption{ The relationship between vortex structures and the work $W^+_{y+z}$.    
		Left column panels are instantaneous isosurfaces of $\lambda_2^+ = -0.007$ colored by the work $W^+_{y+z}$ in half channel and slices of $W^+_{y+z}$ in the y-z and x-z planes  for particle-laden flows with (a) $\lambda =100$; (c) $\lambda =0.01$; (e) $\lambda =0.002$.
		Right column panels are isosurfaces of $\lambda_2^+ = -0.002$ for ensemble-averaged coherent structures with positive streamwise vorticity colored by $W^+_{y+z}$  and slices of $W^+_{y+z}$ in y-z plane for particle-laden flows with (b) $\lambda =100$; (d) $\lambda =0.01$; (f) $\lambda =0.002$.}
	\label{fig13}			
\end{figure}
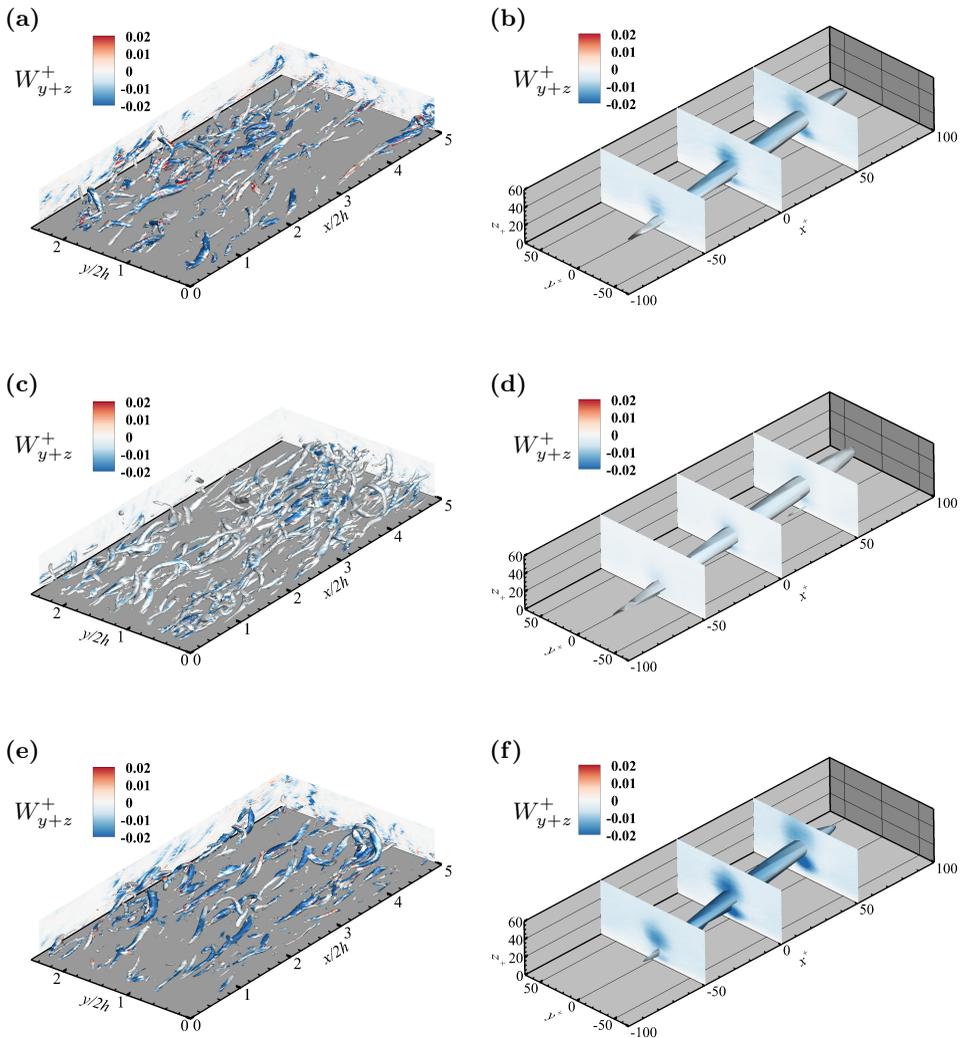

The conditional ensemble average flow fields and the work $W_{y+z}$ are presented to further validate the above findings from a statistical standpoint. The instantaneous quasistreamwise vortices with typical features, such as  $\lambda _2 < -\lambda_{2,\mathrm{rms}}$ , a streamwise length at least 150 wall units, and positive streamwise vorticity,  are sampled conditionally in the range $0 <z^+ < 60$. Then the flow fields and the work $W_{y+z}$ around these structures are ensemble-averaged after proper alignment. It should be noted that  $\lambda_2$ for this flow field is obtained from the ensemble-averaged fluid velocity. The detailed procedures are referred to \cite{jeong_coherent_1997,dritselis_numerical_2008}. As shown in the right column panels of Figure 13, spheroids generally produce negative $W^+_{y+z}$ in the regions of quasistreamwise vortex (see the isosurfaces). The large negative work appears in the upper left and lower right of the vortex (see the slices), where the velocity fluctuations in the spanwise and wall-normal directions are intense. Comparisons between Figure 13 (b), (d) and (f) reveal the dependence of the magnitude of $W^+_{y+z}$ on particle shape. 

Based on these findings, we demonstrate that inertialess spheroids weaken the quasistreamwise vortices through the work $W^+_{y+z}$ whose magnitude depends on particle shape. The Reynolds shear stress in particle-laden flow is then reduced. However, the damping of the Reynolds shear stress $-\langle u'w'\rangle $  cannot directly lead to drag reduction of wall turbulence \citep{eshghinejadfard_lattice_2018}. As shown in Appendix B the drag coefficient $C_f$ can be expressed as:

\begin{equation}
C_f = \frac{\tau_w}{\rho {U_b}^2} 
    =\frac{\nu^2{u_{\tau}}^2}  
    {h^2[\frac{{u_{\tau}}^2}{3}-\frac{1}{\rho}\int_0^1 (1-\eta)(-\rho \langle u'w'\rangle +\langle \tau^p_{xz}\rangle)\mathrm {d}\eta]^2} ,
\label{drag coefficient}
\end{equation}  
where $\eta =z/h$. Because the present flows are driven by a constant mean pressure gradient, the differences in drag coefficients among all cases are only  dependent on Reynolds shear stress and particle induced shear stress. The weighted integration $\bigl( \int_0^1 (1-\eta)(-\rho \langle u'w'\rangle+\langle \tau^p_{xz}\rangle )\mathrm {d}\eta \bigr)$  is positively correlated with the drag coefficient, while negatively with the efficiency of drag reduction. This integration in particle suspensions determines whether drag reduction happens. Figure 14(a) illustrates $-\langle u'w' \rangle $ and $\langle \tau^p_{xz} \rangle $ as functions of the wall distance $z/h$. The particle shear stress compensates for the reduction of Reynolds shear stress, especially in the near-wall region, where the weight factor $(1-\eta)$ is relatively higher. The weighted shear stress $(1-\eta)(-\rho \langle u'w' \rangle +\langle \tau^p_{xz}\rangle )$ is depicted in Figure 14(b). The integrations, i.e. areas under profiles of the weighted shear stresses, are in good agreement with the degree of drag reduction. Compared with the results of rods, the large particle shear stress $\langle \tau^p_{xz} \rangle $ generated by disks in the near-wall region increases the integration and reduces the efficiency of drag reduction. Therefore, it is reasonable that disks are less effective than rods in drag reduction, even though the former also has a significant impact on the turbulence modulations.

\begin{figure}
	\centering
	\begin{tabular}{ccc}	
		\begin{overpic}[width=6.5cm, height=5.0cm
			]{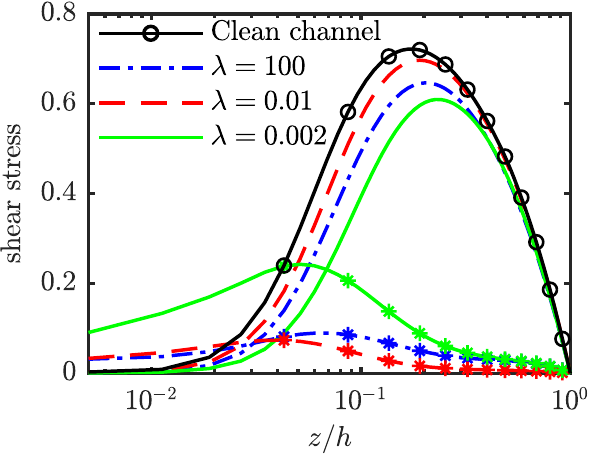}
			\put(0,70){\textbf{(a)}} 
		\end{overpic}  
		&
		
		\begin{overpic}[width=6.5cm,height=5.0cm
			]{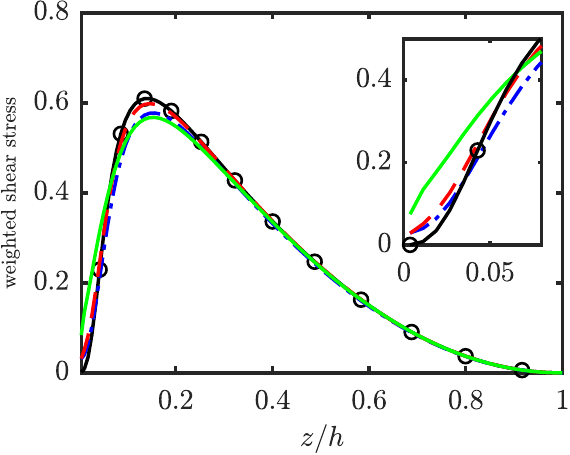}
			\put(0,70){\textbf{(b)}} 
		\end{overpic}    
	\end{tabular}
	\caption{(a) Reynolds shear stress and particle shear stress; (b) the weighted  shear stress. }
	\label{fig14}
\end{figure}

\section{Concluding remarks}\label{conclusions}

The present study has examined the interaction between spheroidal particles and wall turbulence in two-way coupled direct numerical simulations. We focused on how fiber-like and disk-like particles modulate the near-wall turbulence and the mechanism of drag reduction induced by non-spherical particles. Three types of spheroids with aspect ratios of $\lambda =100$, $0.01$ and $0.002$ were considered and 40 million particles of each type, corresponding to a volume fraction of $0.75\%$, were released into fully developed turbulent channel flows at $Re_{\tau}=180$. These tiny inertialess spheroids were tracked in a Lagrangian framework and affected fluid phase via particle stress. One-way coupled simulations, where the feedback from particles onto the fluid was ignored, were also performed for comparison. 

Our results show that rods with $\lambda = 100$ lead to the most pronounced drag reduction ($14.93\%$), followed by disks with $\lambda = 0.002$ ($7.15\%$), and finally disks with $\lambda = 0.01$ ($1.92\%$). We also observed the typical features of drag reduced flows by additives \citep{zhao_turbulence_2010,white_mechanics_2008,paschkewitz_numerical_2004} in particle suspensions. Compared with the unladen flow case, the shear stress balance was modulated and Reynolds shear stress was attenuated. The velocity fluctuations were also altered by the presence of inertialess spheroids that the streamwise velocity fluctuation was enhanced except in the near-wall region, whereas the fluctuations in the other directions were damped. Moreover, the modulated vorticity field indicated that the addition of spheroids resulted in fewer vortices with larger sizes, which was confirmed by the instantaneous vortex structures (Figure 9). Consequently, near-wall streaks became more regular and the mean streak spacing was increased in particle suspensions. On the other hand, the particle dynamics were affected by the modulated fluid field. Similar to the observations in one-way coupled simulations, disk-like particles tended to align norm to the wall, while rod-like particles preferentially aligned parallel to the wall. However, such alignment tendency was strengthened in the two-way coupled simulations. Under the influences of the damped fluid vorticities and  the enhanced alignment tendency, mean angular velocity and spin fluctuations of spheroids were reduced considerably.

The physical mechanism responsible for drag reduction by inertialess spheroids can be depicted as follows. Spheroidal particles weaken the near-wall quasistreamwise vortices through the work $W^+_{y+z}$ whose magnitude depends on particle shape, which reveals that, in the regions of quasistreamwise vortex structures, particle body force tends to suppress the fluid fluctuating motions in the $y$ and $z$ directions. Therefore, the Reynolds shear stress in particle-laden flow is damped, which contributes to the turbulent drag reduction. According to equation (3.3), the degree of drag reduction depends on the profiles of Reynolds shear stress and particle shear stress $\langle \tau^p_{xz} \rangle $. As shown in Figure 14, the particle shear stress partly compensates for the reduction of Reynolds shear stress and limits the efficiency of drag reduction. Because the stress  $\langle \tau^p_{xz} \rangle $ induced by disks is relatively large, especially in the near-wall region, disks induce less pronounced drag reduction than rods. The present results suggest that tiny disk-like particles can lead to a noticeable drag reduction in wall turbulence and can be an alternative drag reduction agent.

\section*{Acknowledgments}
The work was supported by the Natural Science Foundation of China (grant Nos. 11911530141 and 91752205).

\section*{Declaration of Interests}
The authors report no conflict of interest.

\begin{appendices}
	
\section*{Appendix A. Transport equation of turbulent kinetic energy }
\setcounter{equation}{0}
\renewcommand\theequation{A.\arabic{equation}}

By subtracting the mean flow equation from the Navier-Stokes equation (2.2), the fluctuating momentum equation of fluid can be written as:

\begin{equation}
\begin{split}
\rho\left(\frac{\partial u_{i}^{\prime}}{\partial t}+\left\langle u_{j}\right\rangle \frac{\partial u_{i}^{\prime}}{\partial x_{j}}+u_{j}^{\prime} \frac{\partial\left\langle u_{i}\right\rangle}{\partial x_{j}}\right)
&=-\frac{\partial p^{\prime}}{\partial x_{i}}+\mu \frac{\partial^{2} u_{i}^{\prime}}{\partial x_{j} \partial x_{j}}\\
&\quad +\rho \frac{\partial\left\langle u_{i}^{\prime} u_{j}^{\prime}\right\rangle}{\partial x_{j}}-\rho \frac{\partial u_{i}^{\prime} u_{j}^{\prime}}{\partial x_{j}}+\frac{\partial \tau_{i j}^{p^{\prime}}}{\partial x_{j}}.
\end{split}
\end{equation}

Multiplying by $u_i^{\prime}$ and taking the mean of that equation, we obtain

\begin{equation}
\begin{split}
\frac{1}{2}\frac{\partial \langle u_{i}^{\prime} u_{i}^{\prime} \rangle }{\partial t}
+\frac{1}{2}\langle u_{j}\rangle \frac{\partial \langle u_{i}^{\prime} u_{i}^{\prime}\rangle}{\partial x_{j}}
&= -\langle u_{i}^{\prime} u_{j}^{\prime} \rangle \frac{\partial\langle u_{i}\rangle}{\partial x_{j}}\\
&\quad -\frac{\partial}{\partial x_{j}}\left(\frac{\langle p^{\prime} u_{j}^{\prime}\rangle }{\rho}
+\frac{1}{2}\langle u_{i}^{\prime} u_{i}^{\prime}u_{j}^{\prime}\rangle
-\frac{\nu}{2} \frac{\partial \langle u_{i}^{\prime} u_{i}^{\prime} \rangle }{\partial x_{j}}\right)\\
&\quad -\nu\left\langle\frac{\partial u_{i}^{\prime}}{\partial x_{j}} \frac{\partial u_{i}^{\prime}}{\partial x_{j}}\right\rangle\\
&\quad +\frac{1}{\rho}\left\langle u_{i}^{\prime} \frac{\partial \tau_{i j}^{p^{\prime}}}{\partial x_{j}}\right\rangle.
\end{split}
\end{equation}

Then the transport equation of turbulent kinetic energy is given as:

\begin{equation}
\begin{split}
\frac{\partial k }{\partial t} +\langle u_{i}\rangle \frac{\partial k}{\partial x_{i}}
&= -\langle u_{i}^{\prime} u_{j}^{\prime} \rangle \frac{\partial\langle u_{i}\rangle}{\partial x_{j}}\\
&\quad -\frac{\partial}{\partial x_{j}}\left(\frac{\langle p^{\prime} u_{j}^{\prime}\rangle }{\rho}
+\frac{\langle u_{i}^{\prime} u_{i}^{\prime} u_{j}^{\prime}\rangle}{2}
-\nu \frac{\partial k }{\partial x_{j}}\right)\\
&\quad -\nu\left\langle\frac{\partial u_{i}^{\prime}}{\partial x_{j}} \frac{\partial u_{i}^{\prime}}{\partial x_{j}}\right\rangle\\
&\quad +\frac{1}{\rho}\langle u_{i}^{\prime} f_{i}^{\prime}\rangle,
\end{split}
\end{equation}
where $k=\langle u_{i}^{\prime}u_{i}^{\prime}\rangle /2$ is the turbulent kinetic energy and $f_i = {\partial{\tau^{p}_{ij}}}/{\partial{x_j}}$ is defined as particle body force. The last term on the right-hand side is the rate of work done by particles to fluid per mass.

\section*{Appendix B. Drag coefficient of spheroid-laden channel flow}
\setcounter{equation}{0}
\renewcommand\theequation{B.\arabic{equation}}

In order to derive the drag coefficient in the present flow configuration, we start with the mean momentum equation in the streamwise direction:
\begin{equation}
0=-\frac{1}{\rho} \frac{\mathrm{d} \langle p_w\rangle}{\mathrm{d} x}+\nu \frac{\mathrm{d}^{2} U}{\mathrm{d} z^{2}}+\frac{1}{\rho} \frac{\mathrm{d}\langle\tau_{xz}^{p}\rangle}{\mathrm{d} z}-\frac{\mathrm{d} \langle u^{\prime} w^{\prime}\rangle}{\mathrm{d} z},
\end{equation}
where ${\mathrm{d}\langle p_w \rangle}/{\mathrm{d}x}$ is the mean pressure gradient at channel walls in the streamwise direction and $U$ is the mean streamwise velocity. Integrating the above equation from $z$ to $h$ (the half-channel height) in the wall-normal direction, we obtain the shear stress balance equation:
\begin{equation}
\mu \frac{\mathrm{d} U}{\mathrm{d} z}-\rho\left\langle u^{\prime} w^{\prime}\right\rangle+\left\langle\tau_{x z}^{p}\right\rangle=\tau_{w}\left(1-\frac{z}{h}\right).
\end{equation}
Here, $\tau_w=-h{\mathrm{d}\langle p_w \rangle}/{\mathrm{d}x}=\rho u_\tau ^2$ is the wall shear stress. A double integral of this equation from $0$ to $z$ and from $0$ to $h$ in the wall-normal direction gives 
\begin{equation}
h \mu U_{b}+\int_{0}^{h} \int_{0}^{z} \left(-\rho\langle u^{\prime} w^{\prime}\rangle+\langle\tau_{x z}^{p}\rangle\right) \mathrm{d} z \mathrm{d} z=\frac{h^{2}}{3} \tau_{w},
\end{equation}
where the bulk velocity is defined as $U_b=\int_{0}^{h} U \mathrm{d}z/h$. The second term on the right-hand side can be simplified with the application of partial integration:
\begin{equation}
\begin{split}
\int_{0}^{h} \int_{0}^{z} \left(-\rho\langle u^{\prime} w^{\prime}\rangle+\langle\tau_{x z}^{p}\rangle\right) \mathrm{d} z \mathrm{d} z 
& = \left. z\int_{0}^{z} \left(-\rho\langle u^{\prime} w^{\prime}\rangle+\langle\tau_{x z}^{p}\rangle\right)\mathrm{d}z  \right| _{0}^{h} \\
& -\int_{0}^{h} z\left(-\rho\langle u^{\prime} w^{\prime}\rangle+\langle\tau_{x z}^{p}\rangle\right)\mathrm{d}z\\
&=\int_{0}^{h}\left( h-z\right)\left(-\rho\langle u^{\prime} w^{\prime}\rangle+\langle\tau_{x z}^{p}\rangle\right)\mathrm{d}z\\
&=h^2\int_{0}^{1}\left( 1-\eta \right)\left(-\rho\langle u^{\prime} w^{\prime}\rangle+\langle\tau_{x z}^{p}\rangle\right)\mathrm{d}\eta,
\end{split}
\end{equation}
in which $\eta=z/h$ is the dimensionless height from the lower channel wall.
The bulk velocity $U_b$ is then expressed as:
\begin{equation}
U_{b}=\frac{h}{\nu}\left(\frac{u_{\tau}^{2}}{3}-\frac{1}{\rho}\int_{0}^{1}(1-\eta)\left(-\rho\langle u^{\prime} w^{\prime}\rangle+\langle\tau_{x z}^{p}\rangle\right) \mathrm{d} \eta\right).
\end{equation}
Finally, the drag coefficient $C_f$ of the present laden flows is given as follows
\begin{equation}
C_{f}=\frac{\tau_{w}}{\rho U_{b}^{2}}=\frac{\nu ^{2} u_{\tau}^{2}}{h^{2}\left(\frac{u_{r}^{2}}{3}-\frac{1}{\rho} \int_{0}^{1}(1-\eta)\left(-\rho\left\langle u^{\prime} w^{\prime}\right\rangle+\left\langle\tau_{x z}^{p}\right\rangle\right) \mathrm{d} \eta\right)^{2}}.
\end{equation}

\end{appendices}

\bibliographystyle{jfm}
\bibliography{jfm-draft}

\end{document}